\documentclass[dvips,12pt,a4paper]{article}
\usepackage{a4wide}
\usepackage{epsfig}
\usepackage{amsmath}
\usepackage{latexsym}

  \newlength{\absize}
\newcommand{\dd}{\mbox{{\rm d}}}

\newcommand{\Lumint}{{\cal L}_{\rm int}}

\allowdisplaybreaks 

\catcode`@=11
\def\citer{\@ifnextchar [{\@tempswatrue\@citexr}{\@tempswafalse\@citexr[]}}
 
\def\@citexr[#1]#2{\if@filesw\immediate\write\@auxout{\string\citation{#2}}\fi
  \def\@citea{}\@cite{\@for\@citeb:=#2\do
    {\@citea\def\@citea{--\penalty\@m}\@ifundefined
       {b@\@citeb}{{\bf ?}\@warning
       {Citation `\@citeb' on page \thepage \space undefined}}%
\hbox{\csname b@\@citeb\endcsname}}}{#1}}
\catcode`@=12

\begin{document}
  \thispagestyle{empty}
  \pagestyle{empty}
  \renewcommand{\thefootnote}{\fnsymbol{footnote}}
\newpage\normalsize
    \pagestyle{plain}
    \setlength{\baselineskip}{4ex}\par
    \setcounter{footnote}{0}
    \renewcommand{\thefootnote}{\arabic{footnote}}
\newcommand{\preprint}[1]{%
  \begin{flushright}
    \setlength{\baselineskip}{3ex} #1
  \end{flushright}}
\renewcommand{\title}[1]{%
  \begin{center}
    \LARGE #1
  \end{center}\par}
\renewcommand{\author}[1]{%
  \vspace{2ex}
  {\Large
   \begin{center}
     \setlength{\baselineskip}{3ex} #1 \par
   \end{center}}}
\renewcommand{\thanks}[1]{\footnote{#1}}
\renewcommand{\abstract}[1]{%
  \vspace{2ex}
  \normalsize
  \begin{center}
    \centerline{\bf Abstract}\par
    \vspace{2ex}
    \parbox{\absize}{#1\setlength{\baselineskip}{2.5ex}\par}
  \end{center}}

\begin{flushright}
{\setlength{\baselineskip}{2ex}\par
{\tt preprint IC/2002/115}
} 
\end{flushright}
\vspace*{4mm}
\vfill
\title{Bhabha vs. M{\o}ller scattering as a contact-interaction analyzer 
at a polarized Linear Collider}
\vfill
\author{
A. Pankov$^{a,b}$ {\rm and}
N. Paver$^{a}$}
\begin{center}
$^a$ Dipartimento di Fisica Teorica, Universit\`a di Trieste and \\
Istituto Nazionale di Fisica Nucleare, Sezione di Trieste,
Trieste, Italy\\
$^b$ The Abdus Salam International Centre for Theoretical Physics,  Trieste,
Italy
\end{center}
\vfill
\abstract
{We discuss electron-electron contact-interaction searches in the 
processes $e^+e^-\to e^+e^-$ and $e^-e^-\to e^-e^-$ at planned 
Linear Colliders run in the $e^+e^-$ and $e^-e^-$ modes with both beams 
longitudinally polarized. Our analysis is based on the measurement, for the 
two processes, of polarized differential cross sections, and allows to 
simultaneously take into account the general set of electron contact 
interaction couplings as independent, non-zero, parameters thus avoiding the 
simplifying choice of a model. We evaluate the corresponding model-independent 
constraints on the contact coupling constants, emphasizing the role of the 
available beam polarization and the complementarity, as far as the chirality 
of the constants is concerned, of the two processes in giving the best 
constraints. We also make a comparison with the potential of 
$e^+e^-\to\mu^+\mu^-$ at the same energy and initial beams polarization. 
 }
\vspace*{20mm}
\setcounter{footnote}{0}
\vfill

\newpage
    \setcounter{footnote}{0}
    \renewcommand{\thefootnote}{\arabic{footnote}}
    \setcounter{page}{1}

\section{Introduction}
Contact interaction Lagrangians (CI) provide a framework 
to account for the phenomenological effects of non-standard dynamics 
characterized by extremely large intrinsic mass scales $\Lambda$, at the 
`low' energies $\sqrt s\ll\Lambda$ attainable at current particle 
accelerators. One of the historical motivations for considering such a 
framework is the fact that `low energy' manifestations of quark and lepton 
substructure would occur {\it via} four-fermion quark and lepton 
contact interactions, induced by exchanges of quite heavy sub-constituent 
bound states with mass of the order of $\Lambda$. Indeed, in the   
spirit of the `effective interactions', this concept can be used more 
generally, to  parameterize non-standard, very heavy 
particle exchanges in reactions among quarks and leptons, in the form of 
`low energy' expansions of the relevant amplitudes at the leading order in 
$\sqrt s/\Lambda$. Since the above mentioned exchanged heavy particles, 
with mass $M\gg M_{W,Z}$, could not be directly produced at the 
collider energy $\sqrt s$, the underlying 
non-standard dynamics could experimentally manifest itself only indirectly, 
by deviations of the measured observables from the Standard Model (SM) 
predictions. If such deviations were effectively observed to a given 
significance level, one could try to gain numerical information on the 
parameters (masses and coupling constants) of non-standard models and, 
eventually, select the viable ones \cite{perelstein, rizzo}. In the case 
where, instead, no deviation from the SM predictions is observed within the 
experimental accuracy, one can set numerical bounds or constraints on the 
parameters characterizing the new interactions, and determine 
the discovery reach of planned high-energy colliders. Clearly, also this kind 
of information should be  phenomenologically useful in model applications. 
\par 
The explicit form of the contact interaction Lagrangian depends on the kind 
of external particles participating in the considered reaction. For the 
Bhabha scattering process of interest here: 
\begin{equation}
e^++e^-\to e^++e^-,
\label{proc}
\end{equation}
as well as for M{\o}ller scattering:
\begin{equation}
e^-+e^-\to e^-+e^-, \label{proc1}
\end{equation}
we consider the four-fermion contact-interaction 
Lagrangian \cite{Eichten:1983hw}:
\begin{equation}
{\cal L}_{\rm CI}
=\frac{1}{1+\delta_{ef}}\sum_{i,j}g^2_{\rm eff}\hskip
2pt\epsilon_{ij}
\left(\bar e_{i}\gamma_\mu e_{i}\right)
\left(\bar f_{j}\gamma^\mu f_{j}\right).
\label{lagra}
\end{equation}
In Eq.~(\ref{lagra}): $i,j={\rm L,R}$ denote left- or right-handed 
fermion helicities, $\delta_{ef}=1$ for processes (\ref{proc}) 
and (\ref{proc1}) and, if we assumed lepton universality, the 
same Lagrangian, with $\delta_{ef}=0$, is relevant to the annihilation 
processes 
\begin{equation}
e^++e^-\to l^++l^-, 
\label{proc3}
\end{equation}
with $l=\mu,\tau$. The CI coupling constants in Eq.~(\ref{lagra}) are 
parameterized in terms of corresponding mass scales as 
$\epsilon_{ij}={\eta_{ij}}/{{\Lambda^2_{ij}}}$ and, according to the previous 
remarks concerning compositeness, one assumes $g^2_{\rm eff}=4\pi$. Also, 
by convention, one takes $\vert\eta_{ij}\vert=1$ or $\eta_{ij}=0$, 
leaving the energy scales $\Lambda_{ij}$ as free, {\it a priori} independent,  
parameters.
The explicit $SU(3)\times SU(2)\times U(1)$ symmetry of the helicity 
conserving four-fermion lepton contact interaction (\ref{lagra}) reflects 
that the new dynamics are active well-beyond the electroweak scale. 
Furthermore, Eq.~(\ref{lagra}) represents the lowest dimensional operator, 
$D=6$ being the minimum, and higher-dimensional operators, suppressed by 
higher powers of $s/\Lambda^2$, are supposed to be negligible. 
\par 
As anticipated, we will study the effects of the interaction (\ref{lagra}) 
in processes (\ref{proc}) and (\ref{proc1}) at an $e^+e^-$ Linear Collider 
with c.m.\ energy $\sqrt s=0.5\hskip 2pt{\rm TeV}$ and polarized 
electron and positron beams \cite{tdr}. Indeed, the possibility of 
studying $e^-e^-$ initiated processes, in particular new physics, at 
such a facility by turning the positron beam into an 
electron one, has been recently considered with interest  
\cite{santacruz}. Therefore, it should be useful to evaluate,  
and compare, the sensitivities to the CI coupling constants that can 
be obtained from the measurements of processes (\ref{proc}) and 
(\ref{proc1}).    
\par  
Clearly, from current lower bounds on $\Lambda$'s obtained at LEP 
\cite{geweniger, bourilkov}, of the order of 10-15 TeV depending on 
the specific models chosen to fit the data, we can assume $s\ll\Lambda^2$, so 
that the relative size of the deviations from the SM induced by 
Eq.~(\ref{lagra}) is expected to be of order $s/\alpha\Lambda^2$, with 
$\alpha$ the SM coupling (essentially, the fine structure constant), and 
therefore to be quite small.\footnote{For bounds from 
different kinds of processes, in particular on contact
couplings to quarks, see, {\it e.g.}, Refs.~\cite{cheung, zarnecki}.} 
Consequently, very high collider energies and luminosities are required to 
attain a significant sensitivity on these effects. 
\par 
We notice that for the case of the Bhabha process (\ref{proc}), 
Eq.~(\ref{lagra}) envisages the existence of six independent CI models, 
each one contributing to individual helicity amplitudes or combinations of 
them, with {\it a priori} free, and nonvanishing, coefficients 
(basically, $\epsilon_{\rm LL},\epsilon_{\rm RR}\ {\rm and}
\ \epsilon_{\rm LR}=\epsilon_{\rm RL}$ combined with the $\pm$ signs). 
The same is true for the M{\o}ller process (\ref{proc1}).\footnote{In 
general, apart from the $\pm$ possibility, for $e^+e^-\to{\bar f}f$ with 
$f\ne e$ there are four independent CI couplings, so that 
in the present case of processes (\ref{proc}) and (\ref{proc1}) there is 
one free parameter less.} Correspondingly, in principle, a 
model-independent analysis of the data should account for the situation 
where the full Eq.~(\ref{lagra}) is included in the expression for the 
cross section. Potentially, in this case, the different CI couplings may 
interfere and such interference could substantially weaken the bounds 
because, although the different helicity amplitudes by themselves do not 
interfere, the deviations from the SM could be positive for one 
helicity amplitude and negative for another, so that accidental cancellations 
might occur in the sought for deviations of the relevant observables 
from the SM predictions. 
\par 
The analysis of processes (\ref{proc}) and (\ref{proc1}) proposed 
here relies on the initial beams longitudinal polarization envisaged at 
the planned Linear Colliders. The polarization can be exploited to 
extract the values of the 
individual helicity cross sections from suitable combinations of measurable 
polarized cross sections and, consequently, to disentangle the effects 
of the corresponding CI constants $\epsilon_{ij}$, see, {\it e.g.}, 
Ref.~\cite{babich2001}. 
Therefore, all CI couplings of 
Eq.~(\ref{lagra}) are simultaneously included as independent, non vanishing, 
free parameters and, yet, separate constraints (or exclusion 
regions) on their values can be obtained, free from potential weakening 
due to accidental cancellations. In this sense, the procedure should be 
considered as model-independent. We will also make a comparison of the 
results with those obtained from the simplest, model-dependent, procedure of 
assuming non-zero values for only one of the couplings (or one specific 
combination of them) at a time, with all others set to zero. 
\par 
Specifically, in Sect.~2 we introduce the polarized observables 
for the Bhabha and the M{\o}ller processes, Eqs.~(\ref{proc}) and 
(\ref{proc1}), and discuss the sensitivities 
of the different angular ranges to the CI couplings in the two cases. 
In Sect.~3 we perform the numerical analysis, based on a $\chi^2$ procedure, 
to derive the constraints on the CI couplings and establish the 
attainable reach on the mass scales $\Lambda_{ij}$ as a function of the 
integrated luminosity. Sect.~4 contains some conclusive remarks, in particular 
a comparison of the results from the two processes, and with those obtained 
from the annihilation process (\ref{proc3}).   
\section{Polarized observables}
\subsection{Bhabha scattering}
With $P^-$ and $P^+$ the longitudinal polarization  
of the electron and positron beams, respectively, and $\theta$ the angle 
between the incoming and the outgoing electrons in the c.m.\ frame, 
the differential cross section of process (\ref{proc}) at lowest order,
including $\gamma$ and $Z$ exchanges both in the $s$ and $t$ 
channels and the contact interaction (\ref{lagra}), can be written 
in the following form \citer{schrempp,osland}: 
\begin{eqnarray}
\frac{\dd\sigma(P^-,P^+)}{\dd\cos\theta}
&=&\frac{(1+P^-)\,(1-P^+)}4\,\frac{\dd\sigma_{\rm R}}{\dd\cos\theta}+
\frac{(1-P^-)\,(1+P^+)}4\,\frac{\dd\sigma_{\rm L}}{\dd\cos\theta}
\nonumber \\
&+&\frac{1+P^-P^+}2\,\frac{\dd\sigma_{{\rm LR},t}}{\dd\cos\theta}.
\label{cross}
\end{eqnarray}
In Eq.~(\ref{cross}): 
\begin{eqnarray}
\frac{\dd\sigma_{\rm L}}{\dd\cos\theta}&=&
\frac{\dd\sigma_{{\rm LL}}}{\dd\cos\theta}+
\frac{\dd\sigma_{{\rm LR},s}}{\dd\cos\theta},
\nonumber \\
\frac{\dd\sigma_{\rm R}}{\dd\cos\theta}&=&
\frac{\dd\sigma_{{\rm RR}}}{\dd\cos\theta}+
\frac{\dd\sigma_{{\rm RL},s}}{\dd\cos\theta},
\label{sigP}
\end{eqnarray}
with 
\begin{eqnarray}
\frac{\dd\sigma_{{\rm LL}}}{\dd\cos\theta}&=&
\frac{2\pi\alpha^2}{s}\,\big\vert A_{{\rm LL}}\big\vert^2,
\quad
\frac{\dd\sigma_{{\rm RR}}}{\dd\cos\theta}=
\frac{2\pi\alpha^2}{s}\,\big\vert A_{{\rm RR}}\big\vert^2, 
\nonumber \\
\frac{\dd\sigma_{{\rm LR},t}}{\dd\cos\theta}&=&
\frac{2\pi\alpha^2}{s}\,\big\vert A_{{\rm LR},t}\big\vert^2, 
\quad\frac{\dd\sigma_{{\rm LR},s}}{\dd\cos\theta}=
\frac{\dd\sigma_{{\rm RL},s}}{\dd\cos\theta}=
\frac{2\pi\alpha^2}{s}\,\big\vert A_{{\rm LR},s}\big\vert^2, 
\label{helsig}
\end{eqnarray}
and
\begin{eqnarray}
A_{{\rm RR}}&=&\frac{u}{s}\,\left[1+\frac{s}{t}+g_{\rm R}^2\hskip 2pt
\left(\chi_Z(s)+\frac{s}{t}\,\chi_Z(t)\right)+
2\frac{s}{\alpha}\,\epsilon_{\rm RR}\right],
\nonumber \\
A_{{\rm LL}}&=&\frac{u}{s}\,\left[1+\frac{s}{t}+g_{\rm L}^2\hskip 2pt
\left(\chi_Z(s)+\frac{s}{t}\,\chi_Z(t)\right)+
2\frac{s}{\alpha}\,\epsilon_{\rm LL}\right],
\nonumber \\
A_{{\rm LR},s}&=&
\frac{t}{s}\,\left[1+g_{\rm R}\hskip 2pt
g_{\rm L}\,\chi_Z(s)+\frac{s}{\alpha}\,\epsilon_{\rm LR}\right],
\nonumber \\
A_{{\rm LR},t}&=&
\frac{s}{t}\,\left[1+g_{\rm R}\hskip 2pt
g_{\rm L}\chi_Z(t)+\frac{t}{\alpha}\,\epsilon_{\rm LR}\right].
\label{helamp}
\end{eqnarray}
Here: $\alpha$ is the fine structure constant; $t=-s(1-\cos\theta)/2$, 
$u=-s(1+\cos\theta)/2$ and 
$\chi_Z(s)=s/(s-M^2_Z+iM_Z\Gamma_Z)$ and  
$\chi_Z(t)=t/(t-M^2_Z)$ 
represent the $Z$ propagator in the $s$ and $t$ channels, respectively, 
with $M_Z$ and $\Gamma_Z$ the mass and width of the $Z$;  
$g_{\rm R}=\tan\theta_W$, $g_{\rm L}=-\cot{2\,\theta_W}$ are the SM 
right- and left-handed electron couplings of the $Z$, with $\theta_W$ 
the electroweak mixing angle.
\par
With both beams polarized, the polarization of each beam can be changed 
on a pulse by pulse basis. This would allow the separate measurement 
of the polarized cross sections for each of the three polarization 
configurations $++$, $+-$ and $-+$, corresponding to the sets 
of beam polarizations 
$(P^-,P^+)=(P_1,P_2)$, $(P_1,-P_2)$ and $(-P_1,P_2)$, 
respectively, with $P_{1,2}>0$. Specifically, from Eq.~(\ref{cross}), 
with the simplifying notation $\dd\sigma\equiv\dd\sigma/\dd\cos\theta$:
\begin{eqnarray}
{\dd\sigma_{++}}
=\frac{(1+P_1)(1-P_2)}4\,{\dd\sigma_{\rm R}}+\frac{(1-P_1)(1+P_2)}4\,
{\dd\sigma_{\rm L}}&+&\frac{1+P_1P_2}2\,{\dd\sigma_{{\rm LR},t}},
\nonumber \\
{\dd\sigma_{+-}}
=\frac{(1+P_1)(1+P_2)}4\,{\dd\sigma_{\rm R}}+\frac{(1-P_1)(1-P_2)}4\,
{\dd\sigma_{\rm L}}&+&\frac{1-P_1P_2}2\,{\dd\sigma_{{\rm LR},t}},
\nonumber \\
{\dd\sigma_{-+}}
=\frac{(1-P_1)(1-P_2)}4\,{\dd\sigma_{\rm R}}+\frac{(1+P_1)(1+P_2)}4\,
{\dd\sigma_{\rm L}}&+&\frac{1-P_1P_2}2\,{\dd\sigma_{{\rm LR},t}}.
\label{cross4}
\end{eqnarray}
To extract from the measured polarized cross sections the values of 
$\dd\sigma_{\rm R}$, $\dd\sigma_{\rm L}$ and $\dd\sigma_{{\rm LR},t}$,  
that carry the information on the CI couplings, one has to invert the 
system of equations (\ref{cross4}). The 
solution reads:
\begin{eqnarray}
{\dd\sigma_{\rm R}}&=&
\frac{(1+P_2)^2}{2P_2(P_1+P_2)}\,{\dd\sigma_{+-}}+
\frac{(1-P_1)^2}{2P_1(P_1+P_2)}\,{\dd\sigma_{-+}}-
\frac{1-P_1P_2}{2P_1P_2}\,{\dd\sigma_{++}},
\nonumber \\
{\dd\sigma_{\rm L}}&=&
\frac{(1-P_2)^2}{2P_2(P_1+P_2)}\,{\dd\sigma_{+-}}+
\frac{(1+P_1)^2}{2P_1(P_1+P_2)}\,{\dd\sigma_{-+}}-
\frac{1-P_1P_2}{2P_1P_2}\,{\dd\sigma_{++}},
\nonumber \\
{\dd\sigma_{{\rm LR},t}}&=&
-\frac{1-P_2^2}{2P_2(P_1+P_2)}\,{\dd\sigma_{+-}}-
\frac{1-P_1^2}{2P_1(P_1+P_2)}\,{\dd\sigma_{-+}}+
\frac{1+P_1P_2}{2P_1P_2}\,{\dd\sigma_{++}}.
\label{observ}
\end{eqnarray}
As one can see from Eqs.(\ref{sigP})-(\ref{helamp}), $\sigma_{{\rm LR},t}$ 
depends on a single contact interaction parameter ($\epsilon_{\rm LR}$), 
while $\sigma_{\rm R}$ and $\sigma_{\rm L}$ depend on pairs of parameters,  
($\epsilon_{\rm RR}$,$\epsilon_{\rm LR}$) and 
($\epsilon_{\rm LL}$,$\epsilon_{\rm LR}$), respectively. Therefore, the 
derivation of the model-independent constraints on the CI couplings requires 
the combination of all polarized cross sections as in Eq.~(\ref{observ}). 
In this regard, to emphasize the role of polarization, one can easily 
notice from Eqs.~(\ref{cross})-(\ref{helamp}) that in the unpolarized case 
$P_1=P_2=0$, the interference of the 
$\epsilon_{\rm LR}$ term with the SM amplitude in 
$A_{{\rm LR}s}$ and $A_{{\rm LR},t}$ has opposite signs, leading to a partial 
cancellation for $-t\sim s$. Consequently, as briefly anticipated in Sect.~1, 
one can expect the unpolarized cross section to have reduced sensitivity to 
$\epsilon_{\rm LR}$. Conversely, $\epsilon_{\rm LR}$ is {\it directly} 
accessible from $\dd\sigma_{{\rm LR},t}$, {\it via} polarized cross sections as 
in Eq.~(\ref{observ}). Also, considering that numerically 
$g_{\rm L}^2\cong g_{\rm R}^2$, the parameters $\epsilon_{\rm LL}$ and 
$\epsilon_{\rm RR}$ contribute to the unpolarized cross section through 
$A_{{\rm RR}}$ and $A_{{\rm LL}}$ with equal coefficients, so that, in general, only correlations of the 
form $\vert\epsilon_{\rm LL}+\epsilon_{\rm RR}\vert<{\rm const}$, 
and not finite allowed regions, could be derived in the unpolarized case. 
\par  
To make contact to the experimental situation we take $P_1=0.8$ and $P_2=0.6$, 
and impose a cut in the forward and backward directions. Specifically, 
we consider the cut angular range $\vert\cos\theta\vert<0.9$ and divide 
it into nine equal-size bins of width $\Delta z=0.2$ ($z\equiv\cos\theta$). 
We also introduce the experimental efficiency, $\epsilon$, for
detecting the final $e^+e^-$ pair and $\epsilon\simeq 100\%$ is assumed. 
\par 
We then define the three, directly measurable, event rates integrated 
over each bin:
\begin{equation}
N_{++},\quad  N_{+-},\quad N_{-+},
\label{obsn}
\end{equation}
and ($\alpha\beta=++$, etc.):
\begin{equation}
N_{\alpha\beta}^{\rm bin}=\frac{1}{3}\Lumint{(e^+e^-)}\,\epsilon
\int_{\rm bin}(\dd\sigma_{\alpha\beta}/\dd z)\dd z.
\label{n}
\end{equation}
In Eq.~(\ref{n}), $\Lumint$ is the time-integrated luminosity, which is 
assumed to be equally divided among the three combinations of 
electron and positron beam polarizations defined in Eqs.~(\ref{cross4}). 
\par 
In Fig.~1, the bin-integrated angular distributions of $N_{++}^{\rm bin}$ 
\begin{figure}[thb]
\refstepcounter{figure}
\label{Fig1}
\addtocounter{figure}{-1}
\begin{center}
\setlength{\unitlength}{1cm}
\begin{picture}(8,8)
\put(-1.0,-1.5)
{\mbox{\epsfysize=10.0cm\epsffile{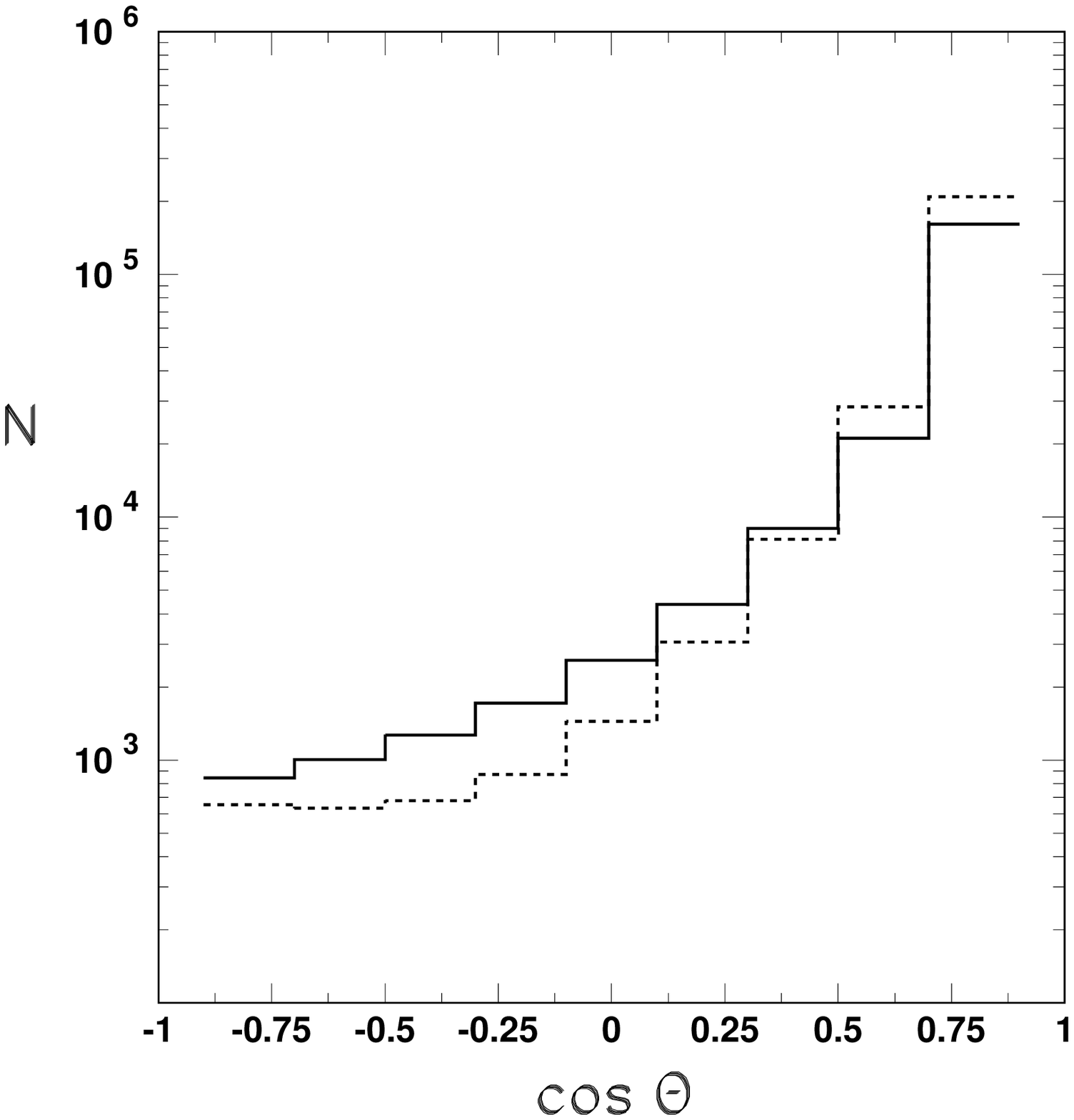}}}
\end{picture}
\vspace*{5mm}
\caption{Bin-integrated angular distributions of
$N_{++}^{\rm bin}$ (solid line) and $N_{+-}^{\rm bin}$ (dashed line), 
Eq.(\ref{n}), in the SM at $\sqrt{s}=500$ GeV,   
$\Lumint(e^+e^-)=50\ \mbox{fb}^{-1}$, $\vert P^-\vert=0.8$ and 
$\vert P^+\vert=0.6$.
}
\end{center}
\end{figure}
and $N_{+-}^{\rm bin}$ in the SM at $\sqrt{s}=500$ GeV and  
$\Lumint=50\ \mbox{fb}^{-1}$ are presented as histograms. Here, the SM 
cross sections have been evaluated by means of the effective 
Born approximation \cite{consoli, altarelli}.
The typical forward peak, dominated by the $t$-channel photon pole,  
dramatically shows up, and determines a really large statistics 
available in the region of small $t$. The $\cos\theta$ distribution for 
the remaining polarization configuration $N_{-+}^{\rm bin}$ in 
(\ref{cross4}) is similar and, therefore, we do not represent it here. 
\begin{figure}[b]
\refstepcounter{figure}
\label{Fig2}
\addtocounter{figure}{-1}
\begin{center}
\setlength{\unitlength}{1cm}
\begin{picture}(12.,6.)
\put(-4.,0.0)
{\mbox{\epsfysize=6.4cm\epsffile{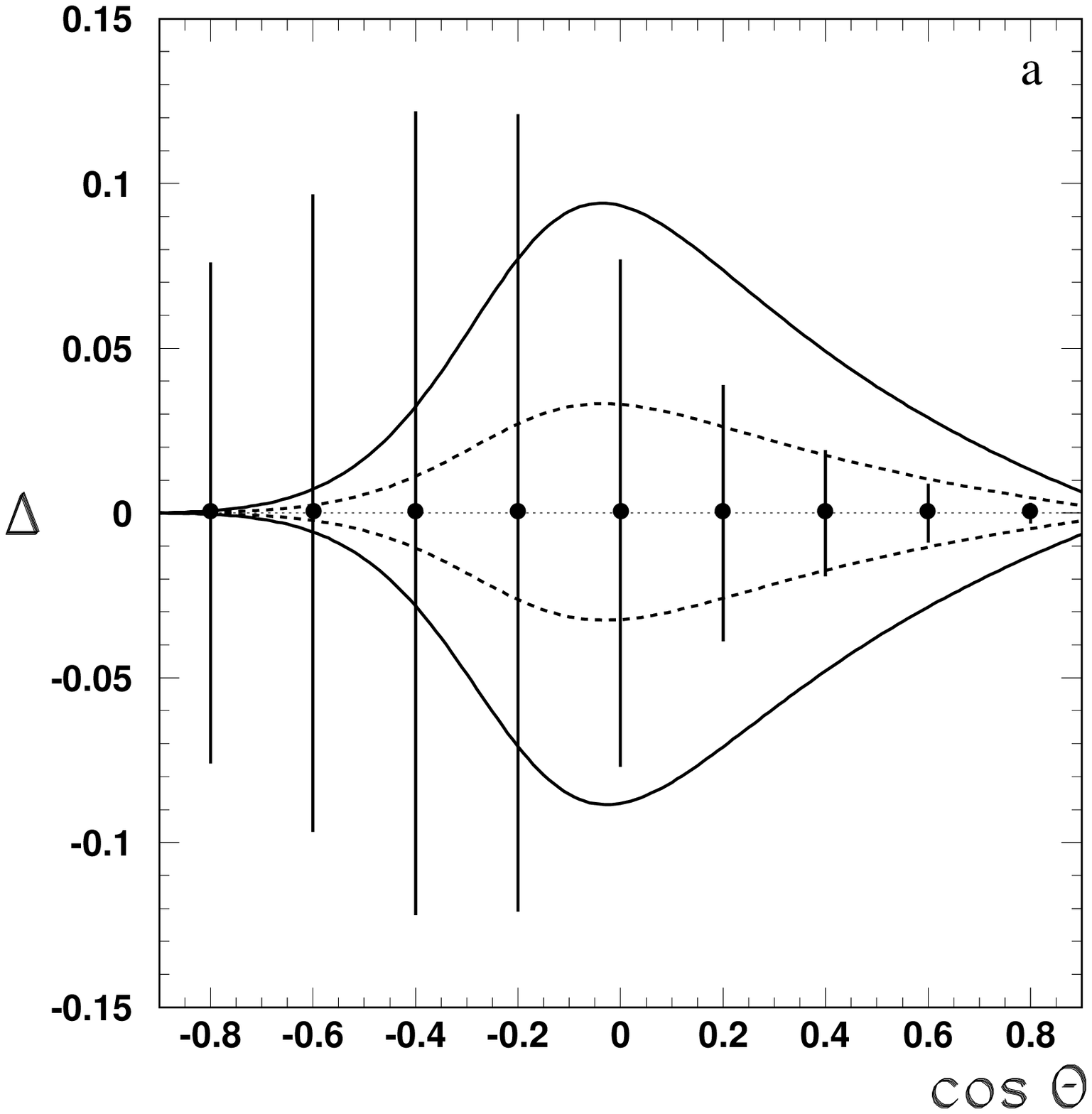}}
 \mbox{\epsfysize=6.4cm\epsffile{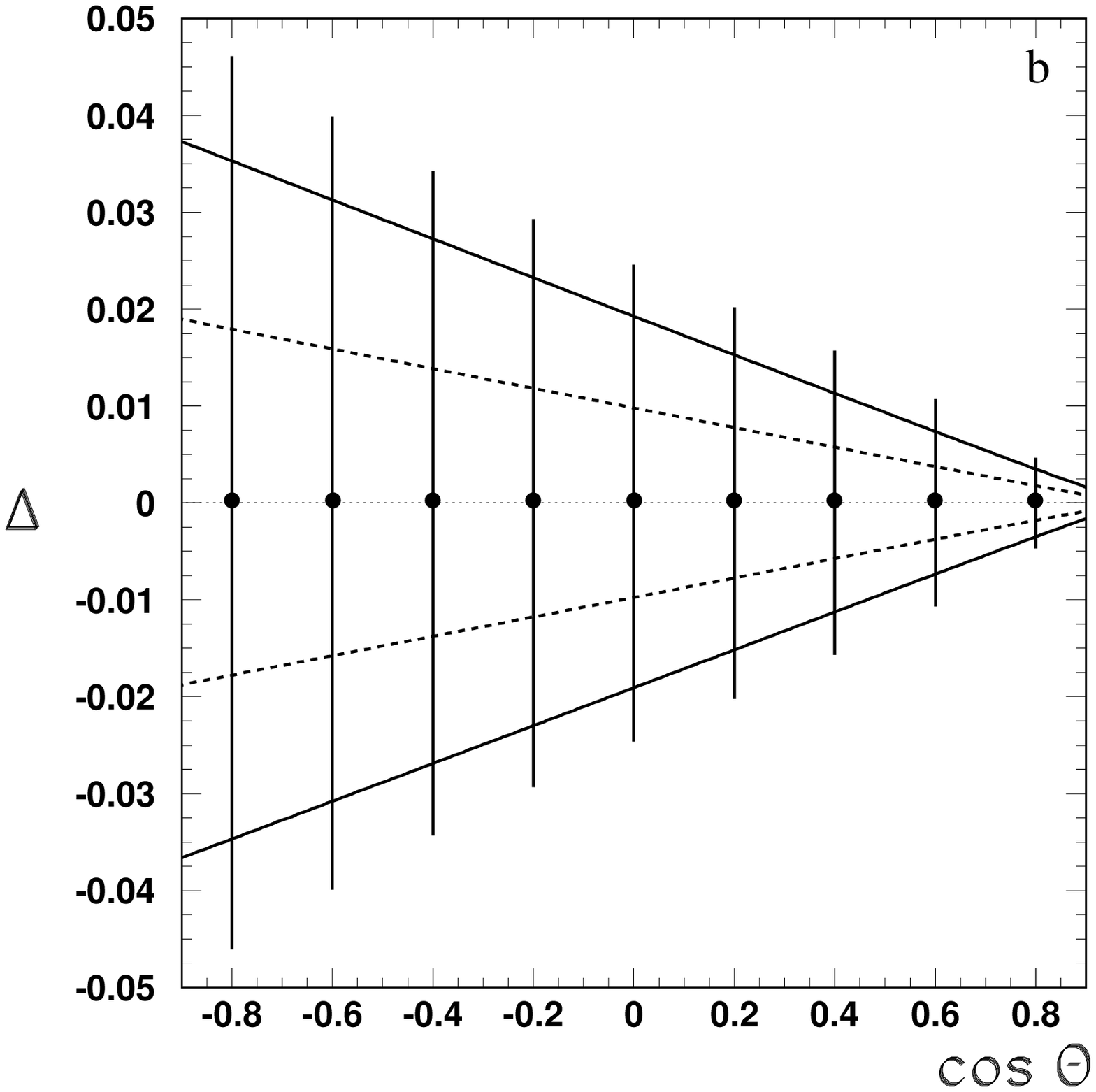}}
 \mbox{\epsfysize=6.4cm\epsffile{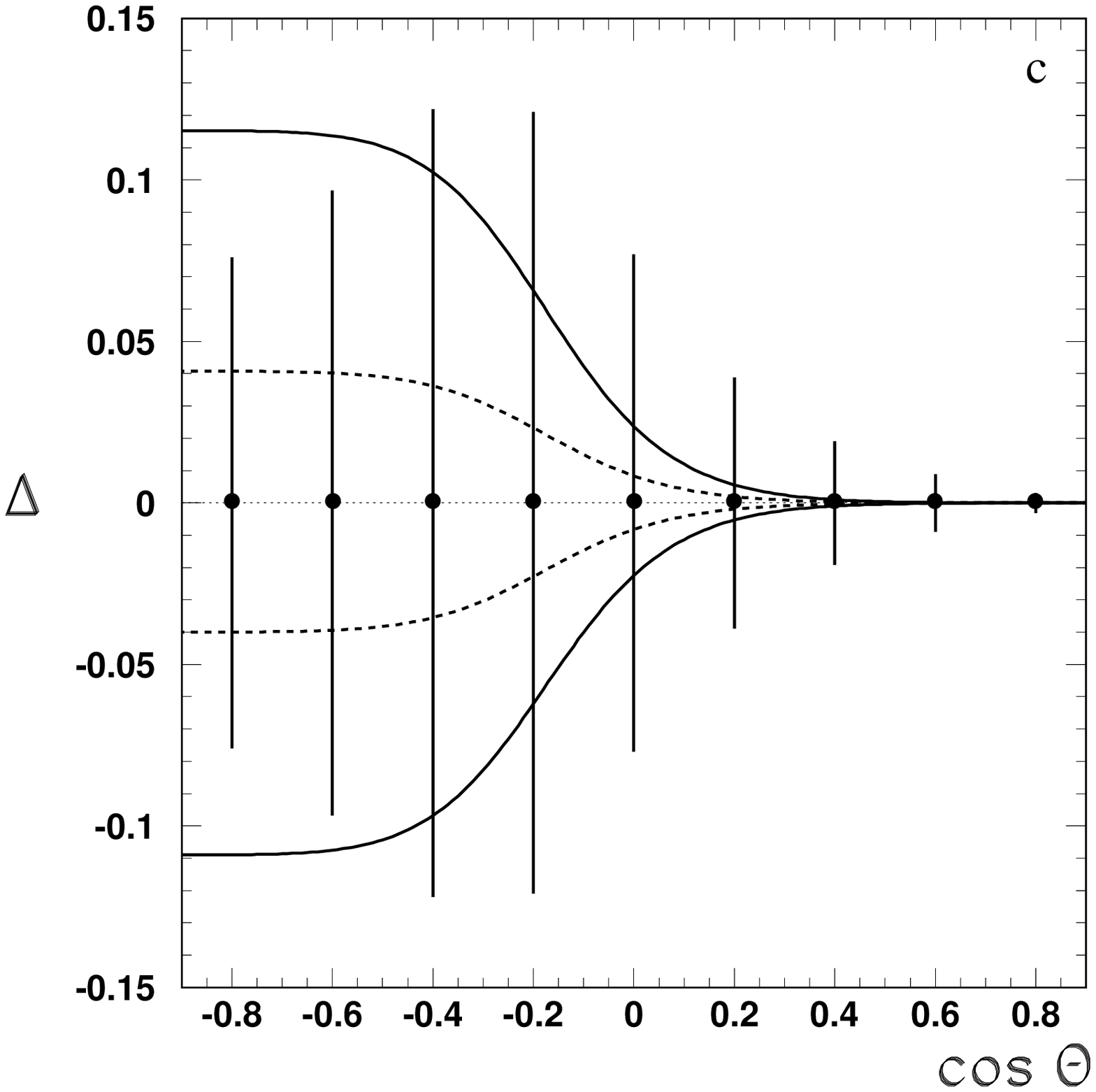}}}
\end{picture}
\vspace*{-3mm}
\caption{The angular distributions of relative deviations from the SM 
predictions: (a) $\Delta(\sigma_{\rm R})$ for $\Lambda_{\rm RR}$=30 TeV 
(solid line) and 50 TeV (dashed line); (b) $\Delta(\sigma_{{\rm LR},t})$
for $\Lambda_{\rm LR}$=50 TeV (solid line) and 70 TeV (dashed line); (c) 
$\Delta(\sigma_{\rm R})$ for $\Lambda_{\rm LR}$=30 TeV (solid line) and
50 TeV (dashed line). In (a) and (b) the curves above (below) the horizontal
line correspond to negative (positive) interference between contact
interaction and SM amplitude, whereas the opposite occurs in (c). The error 
bars show the expected statistical error at $\Lumint(e^+e^-)=50\
\mbox{fb}^{-1}$.}
\end{center}
\end{figure}
\begin{figure}[htb]
\refstepcounter{figure}
\label{Fig3}
\addtocounter{figure}{-1}
\begin{center}
\setlength{\unitlength}{1cm}
\begin{picture}(12,7.5)
\put(-3.,0.0)
{\mbox{\epsfysize=8.5cm\epsffile{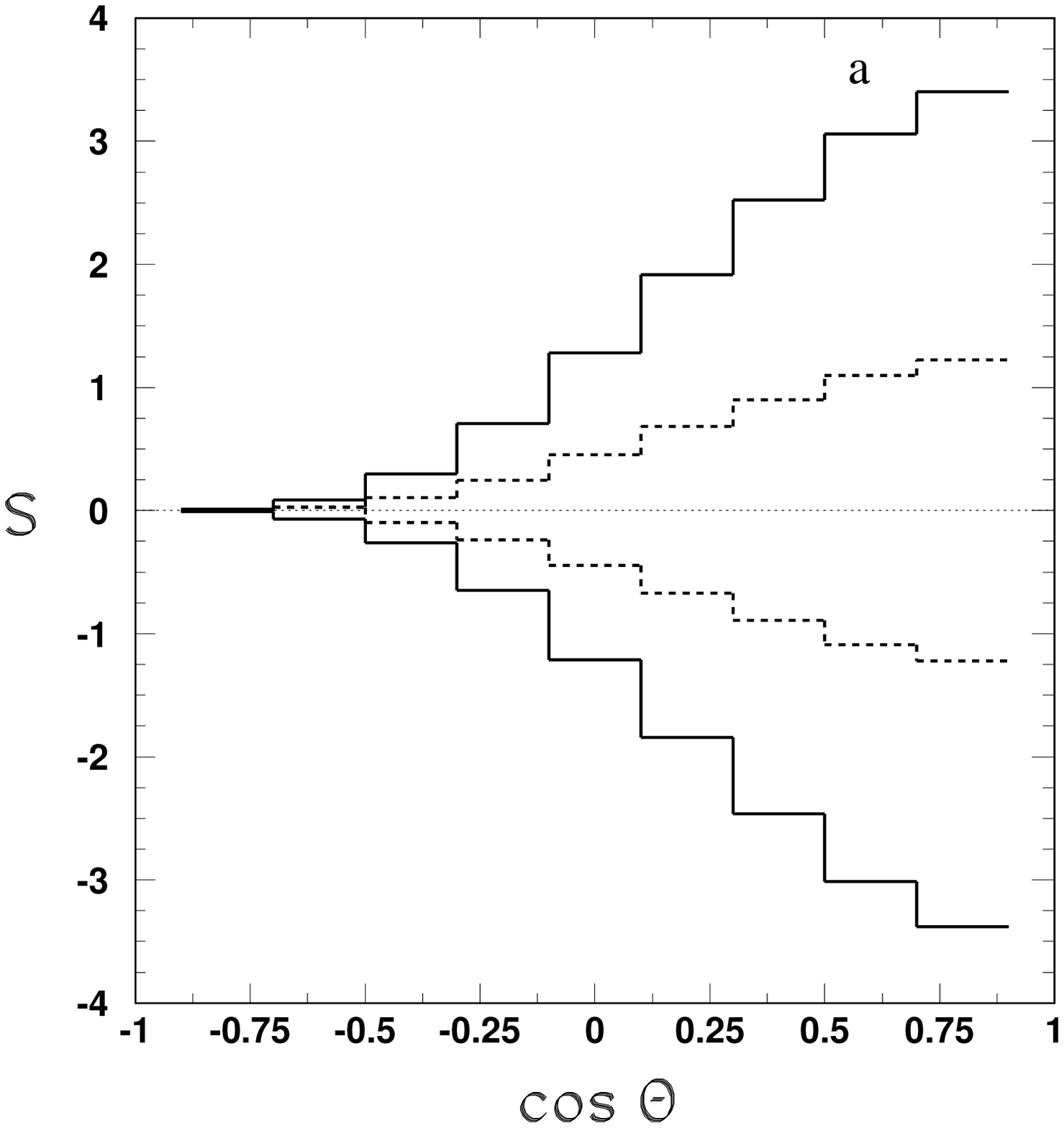}}
 \mbox{\epsfysize=8.5cm\epsffile{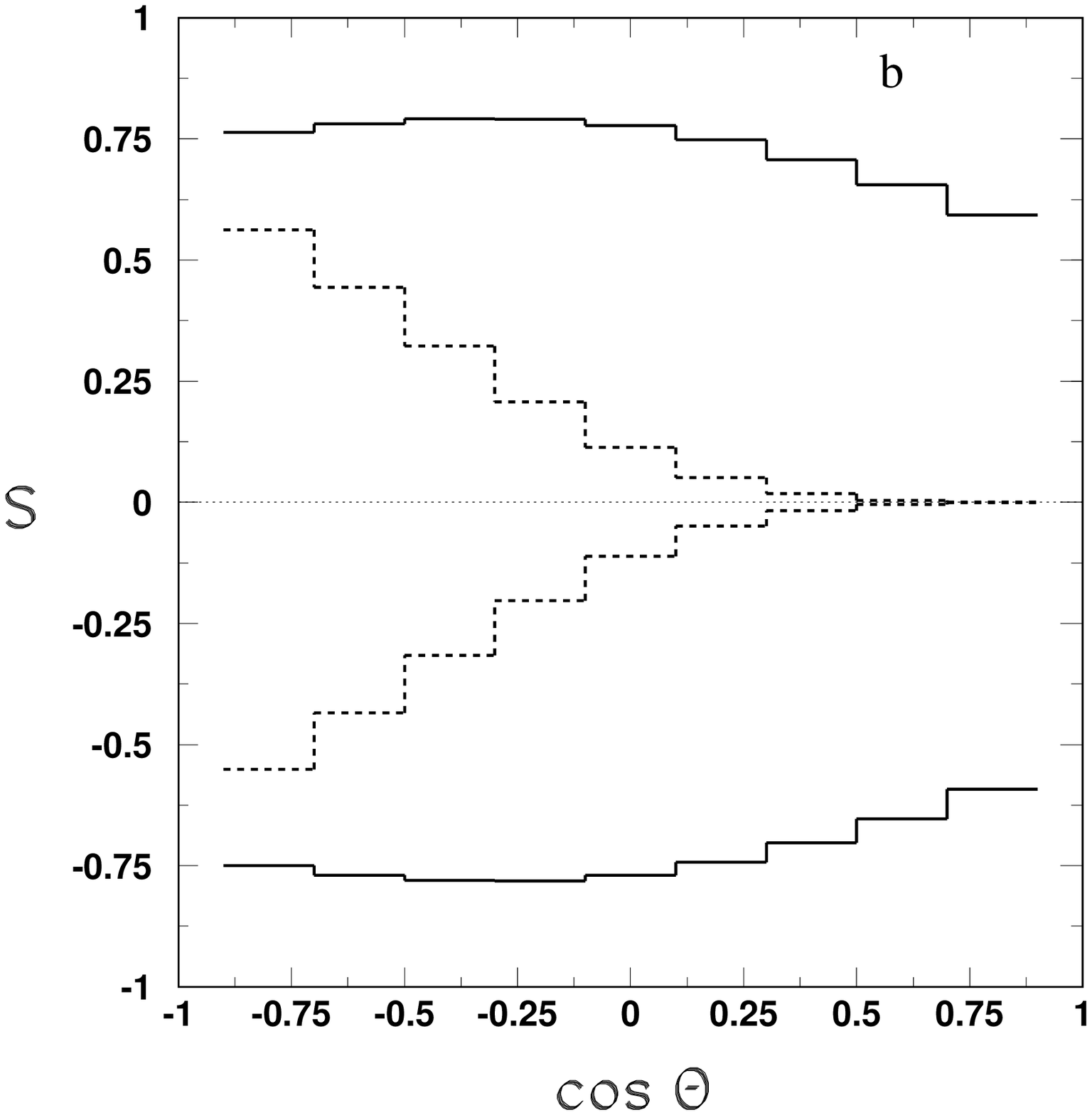}}}
\end{picture}
\vspace*{-5mm}
\caption{
(a) Statistical significance ${\cal S}(\sigma_R)$ as a function 
of $\cos\theta$ at 
$\Lambda_{\rm RR}$=30 TeV (solid line) and 50 TeV (dashed line); 
(b) Statistical significance ${\cal S}(\sigma_R)$ (dashed line) and 
${\cal S}(\sigma_{LR,t})$ (solid line)
as a function of $\cos\theta$ at 
$\Lambda_{\rm LR}$=50 TeV.
Here: $\sqrt{s}=500$ GeV, $\Lumint(e^+e^-)=50\ \mbox{fb}^{-1}$,
$\vert P^-\vert=0.8$ and $\vert P^+\vert=0.6$
}
\end{center}
\end{figure}
\par 
The next step is to define the relative deviations of the polarized 
cross sections from the SM predictions, due to the contact interaction. 
In general, for such deviations, we use the notation: 
\begin{equation}
\Delta ({\cal O})=\frac{{\cal O}(SM+CI)-{\cal O}(SM)}{{\cal O}(SM)},
\label{relat}
\end{equation}  
where 
${\cal O}=\sigma_{\rm R},\, \sigma_{\rm L}\, {\rm and}\, \sigma_{{\rm LR},t}$. 
To get an illustration of the effect of the contact interactions on the 
observables (\ref{observ}) under consideration, we show in 
Figs.~2a,b,c the angular distributions of the relative deviations of
$\dd\sigma_{\rm R}$ and $\dd\sigma_{{\rm LR},t}$, taking as examples the 
values of $\Lumint$ and $\Lambda_{ij}$ indicated in the caption. 
The SM predictions are evaluated in the same effective Born approximation 
as in Fig.~1. The deviations $\Delta ({\cal O})$ are then compared to the 
expected statistical relative uncertainties, represented by the vertical bars. 
Figs.~2a,c show that $\dd\sigma_{\rm R}$ is sensitive to the contact 
interaction $\epsilon_{\rm RR}$ in the forward region, where the ratio of 
the `signal' to the statistical uncertainty substantially increases, while it 
is sensitive to $\epsilon_{\rm LR}$ in the backward 
direction. Also, it qualitatively indicates that, for the chosen values of 
the c.m. energy $\sqrt s$ and $\Lumint$, the reach on $\Lambda_{\rm RR}$ 
will be substantially larger than 30 TeV. Conversely, Fig.~2b shows that the 
sensitivity of $\dd\sigma_{{\rm LR},t}$ is almost independent on the chosen 
kinematical range in $\cos\theta$, leading to a really high sensitivity of 
this observable to $\epsilon_{\rm LR}$, and to a corresponding reach 
on $\Lambda_{\rm LR}$ potentially larger than 50 TeV. The corresponding 
behaviour of the statistical significances, defined as the ratio between the 
deviation from the SM and the statistical uncertainty for each bin, 
${\cal S}({\cal O})=\Delta({\cal O})/\delta{\cal O}$ with 
$\delta{\cal O}$ the expected statistical relative uncertainty, are shown 
in Figs.~3a,b.\footnote{One can notice from Eq.~(\ref{helamp}) that the 
statistical significance $\cal S$ goes to zero in the limit $\theta\to 0$. 
This is not evident from Figs.~3a,b due to the limited kinematical region 
$\vert\cos\theta\vert<0.9$ taken in our analysis.}  
\subsection{M{\o}ller scattering}
With $P_1^-$ and $P_2^-$ the longitudinal polarization  
of the electron beams, 
the differential cross section of process (\ref{proc1}) can be written 
in the following form \citer{cuypers1,marciano}:\footnote{In the case 
of M{\o}ller scattering one can find for the cross section results similar to 
Bhabha scattering, that can be obtained by crossing symmetry except for 
the overall normalization factor 1/2 related to identical particles.} 
\begin{eqnarray}
\frac{\dd\sigma(P_1^-,P_2^-)}{\dd\cos\theta}
&=&\frac{(1+P_1^-)\,(1+P_2^-)}4\,\frac{\dd\sigma_{\rm RR}}{\dd\cos\theta}+
\frac{(1-P_1^-)\,(1-P_2^-)}4\,\frac{\dd\sigma_{\rm LL}}{\dd\cos\theta}
\nonumber \\
&+&\frac{1-P_1^-P_2^-}2\,
\left(\frac{\dd\sigma_{{\rm LR},t}}{\dd\cos\theta}+
\frac{\dd\sigma_{LR,u}}{\dd\cos\theta}\right).
\label{crossmol}
\end{eqnarray}
\newpage
In Eq.~(\ref{crossmol}):
\begin{eqnarray}
\frac{\dd\sigma_{\rm RR}}{\dd\cos\theta}&=&
\frac{\pi\alpha^2}{s}\big\vert A_{\rm RR}\big\vert^2,
\quad
\frac{\dd\sigma_{\rm LL}}{\dd\cos\theta}=
\frac{\pi\alpha^2}{s}\big\vert A_{\rm LL}\big\vert^2,
\nonumber \\
\frac{\dd\sigma_{{\rm LR},u}}{\dd\cos\theta}&=&
\frac{\pi\alpha^2}{s}\big\vert A_{{\rm LR},u}\big\vert^2,
\quad
\frac{\dd\sigma_{{\rm LR},t}}{\dd\cos\theta}=
\frac{\pi\alpha^2}{s}\big\vert A_{{\rm LR},t}\big\vert^2,
\label{helsigmol}
\end{eqnarray}
and
\begin{eqnarray}
A_{\rm RR}&=&\frac{s}{t}\left[1+\frac{t}{u}+g_{\rm R}^2\hskip 2pt
\left(\chi_Z(t)+ \frac{t}{u}\,\chi_Z(u)\right)+
2\frac{t}{\alpha}\,\epsilon_{\rm RR}\right],
\nonumber \\
A_{\rm LL}&=&\frac{s}{t}\left[1+\frac{t}{u}+g_{\rm L}^2\hskip 2pt
\left(\chi_Z(t)+ \frac{t}{u}\,\chi_Z(u)\right)+
2\frac{t}{\alpha}\,\epsilon_{\rm LL}\right],
\nonumber \\
A_{{\rm LR},u}&=&\frac{t}{u}\left[1+g_{\rm R}\hskip 2pt
g_{\rm L}\,\chi_Z(u)+\frac{u}{\alpha}\,\epsilon_{\rm LR}\right],
\nonumber \\
A_{{\rm LR},t}&=&\frac{u}{t}\left[1+g_{\rm R}\hskip 2pt
g_{\rm L}\,\chi_Z(t)+\frac{t}{\alpha}\,\epsilon_{\rm LR}
\right],
\label{helampmol}
\end{eqnarray}
where $\chi_Z(u)=u/(u-M^2_Z)$. Notice that the amplitudes $A_{ij}$ are 
now functions of $t$ and $u$ instead of $t$ and $s$ as in the case of 
Bhabha scattering.
\par
As for the previous process, with both beams polarized the polarization of 
each electron beam can be changed on a pulse by pulse basis. This would 
allow the separate measurement of the polarized cross sections for each of 
the three polarization configurations $++$, $--$ and $+-$, corresponding to 
the sets of beam polarizations $(P_1^-,P_2^-)=(P_1,P_2)$, $(-P_1,-P_2)$ 
and $(P_1,-P_2)$, respectively, with $P_{1,2}>0$. From Eq.~(\ref{crossmol}): 
\begin{eqnarray}
{\dd\sigma_{++}}&=&
\frac{(1+P_1)(1+P_2)}4\,{\dd\sigma_{\rm RR}}+\frac{(1-P_1)(1-P_2)}4\,
{\dd\sigma_{\rm LL}}+\frac{1-P_1P_2}2\,{\dd\sigma_{\rm LR}},
\nonumber \\
{\dd\sigma_{--}}&=& 
\frac{(1-P_1)(1-P_2)}4\,{\dd\sigma_{\rm RR}}+\frac{(1+P_1)(1+P_2)}4\,
{\dd\sigma_{\rm LL}}+\frac{1-P_1P_2}2\,{\dd\sigma_{\rm LR}},
\nonumber \\
{\dd\sigma_{+-}}&=&
\frac{(1+P_1)(1-P_2)}4\,{\dd\sigma_{\rm RR}}+\frac{(1-P_1)(1+P_2)}4\,
{\dd\sigma_{\rm LL}}+\frac{1+P_1P_2}2\,{\dd\sigma_{\rm LR}}.
\label{cross4mol}
\end{eqnarray}
To extract from the measured polarized cross sections the values of 
$\dd\sigma_{\rm RR}$, $\dd\sigma_{\rm LL}$ and $\dd\sigma_{\rm LR}$,  
that carry the information on individual CI couplings, one has to invert the 
system of equations (\ref{cross4mol}). The 
solution reads:
\begin{eqnarray}
{\dd\sigma_{\rm RR}}&=&
\frac{(1+P_2)^2}{2P_2(P_1+P_2)}\,{\dd\sigma_{++}}+
\frac{(1-P_1)^2}{2P_1(P_1+P_2)}\,{\dd\sigma_{--}}-
\frac{1-P_1P_2}{2P_1P_2}\,{\dd\sigma_{+-}},
\nonumber \\
{\dd\sigma_{\rm LL}}&=&
\frac{(1-P_2)^2}{2P_2(P_1+P_2)}\,{\dd\sigma_{++}}+
\frac{(1+P_1)^2}{2P_1(P_1+P_2)}\,{\dd\sigma_{--}}-
\frac{1-P_1P_2}{2P_1P_2}\,{\dd\sigma_{+-}},
\nonumber \\
{\dd\sigma_{\rm LR}}&=&
-\frac{1-P_2^2}{2P_2(P_1+P_2)}\,{\dd\sigma_{++}}-
\frac{1-P_1^2}{2P_1(P_1+P_2)}\,{\dd\sigma_{--}}+
\frac{1+P_1P_2}{2P_1P_2}\,{\dd\sigma_{+-}}.
\label{observmol}
\end{eqnarray}
As one can see from Eqs.~(\ref{helsigmol}) and (\ref{helampmol}), 
contrary to the case of Bhabha scattering, each of the cross sections 
$\sigma_{\rm RR}$, $\sigma_{\rm LL}$ {\it and} $\sigma_{\rm LR}$ depend 
on an individual contact interaction parameter, so that 
full disentanglement of the various couplings (hence the derivation of 
model-independent constraints) is directly obtained by electron beams 
polarization in the M{\o}ller process. 
\par
Similar to Sect.~2.1, see Eqs.~(\ref{obsn}) and (\ref{n}), we define 
measurable event rates integrated over each bin in $z=\cos\theta$:   
\begin{equation}
N_{++},\quad  N_{--},\quad N_{+-},
\label{obsnmol}
\end{equation}
and ($\alpha\beta=++$, etc.):
\begin{equation}
N_{\alpha\beta}^{\rm bin}=\frac{1}{3}\Lumint{(e^-e^-)}\,\epsilon
\int_{\rm bin}(\dd\sigma_{\alpha\beta}/\dd z)\dd z.
\label{nmol}
\end{equation}
In Eq.~(\ref{nmol}), $\Lumint$ is the time-integrated luminosity in the 
$e^-e^-$ mode of the Linear Collider, and is assumed to be equally 
divided among the three combinations of electron beams polarizations 
defined in (\ref{cross4mol}). To account for the lower luminosity in the 
$e^-e^-$ mode due to anti-pinching in the interaction region 
\cite{santacruz, spencer}, we assume 
$\Lumint (e^-e^-)\simeq\frac{1}{3}\Lumint (e^+e^-)$. Also, as regards 
the longitudinal polarization of electrons, we take the symmetric 
configuration $\vert P_1^-\vert=\vert P_2^-\vert=0.8$. 
\par 
Fig.~4 is the analogue of Fig.~1 for Bhabha scattering and represents the 
\begin{figure}[thb]
\refstepcounter{figure}
\label{Fig4}
\addtocounter{figure}{-1}
\begin{center}
\setlength{\unitlength}{1cm}
\begin{picture}(8.,8.)
\put(-1.0,-1.5)
{\mbox{\epsfysize=10.0cm\epsffile{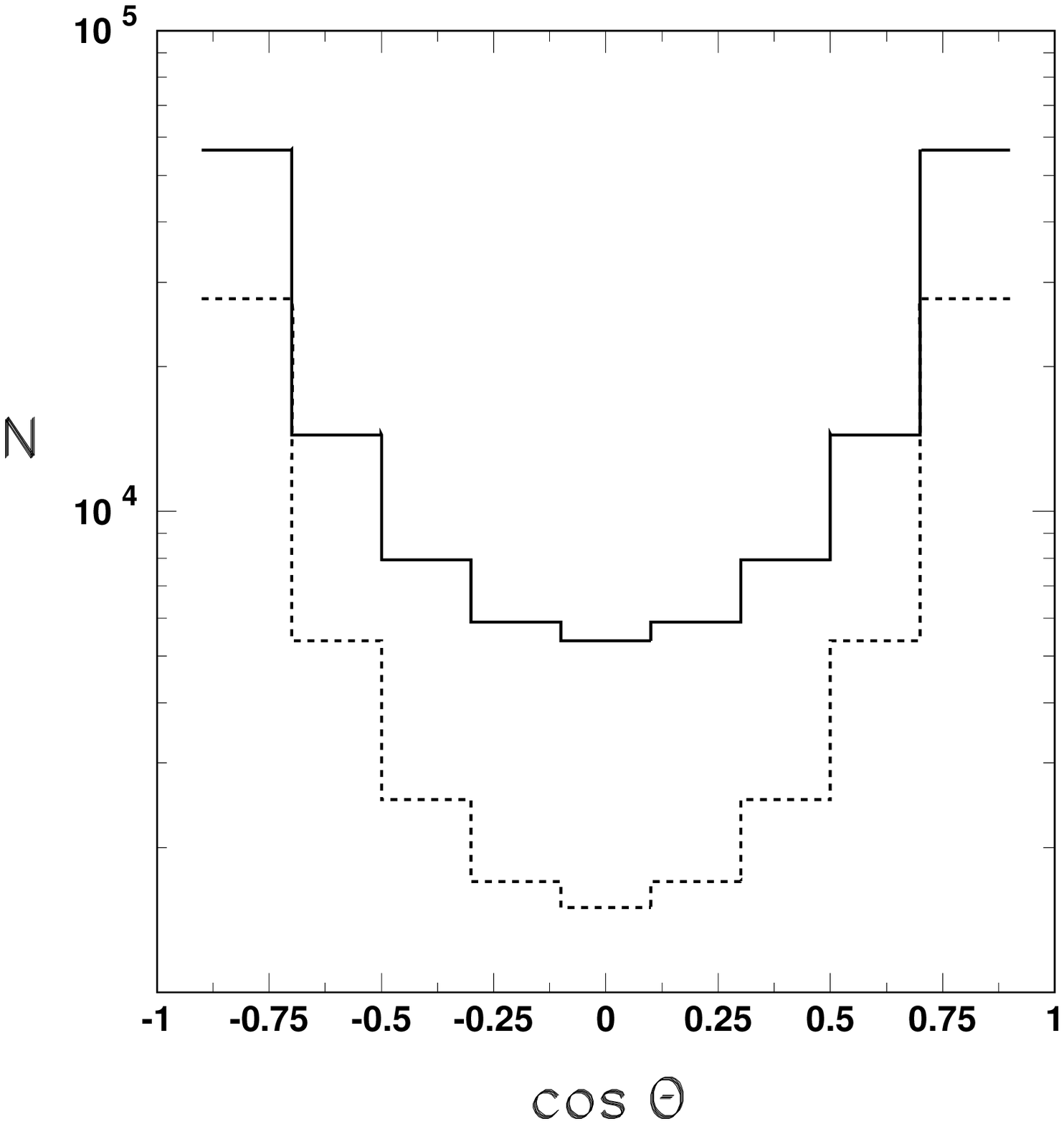}}}
\end{picture}
\vspace*{7mm}
\caption{Bin-integrated angular distributions of
$N_{++}^{\rm bin}$ (solid line) and $N_{+-}^{\rm bin}$ (dashed line) 
in the SM at $\sqrt{s}=500$ GeV, 
$\Lumint(e^-e^-)=\Lumint(e^+e^-)/3$ with 
$\Lumint{(e^+e^-)}=50\, {\rm fb}^{-1}$ and 
$\vert P_1^-\vert=\vert P_2^-\vert=0.8$.
}
\end{center}
\end{figure}
bin-integrated angular distributions of $N_{++}^{\rm bin}$ and 
$N_{+-}^{\rm bin}$ in the SM, calculated by means of the effective Born 
approximation, for the c.m. energy and integrated $e^--e^-$ luminosity 
indicated in the caption. One should notice, in this case, the peaks in 
the forward and backward directions, dominated by the $t$ and $u$ photon 
poles leading to high statistics in those kinematical regions, and the 
dip at $90^\circ$. The $\cos\theta$ distribution for $N_{--}^{\rm bin}$ 
has similar features. 
\begin{figure}[htb]
\refstepcounter{figure}
\label{Fig5}
\addtocounter{figure}{-1}
\begin{center}
\setlength{\unitlength}{1cm}
\begin{picture}(12,7.5)
\put(-3.,0.0)
{\mbox{\epsfysize=8.5cm\epsffile{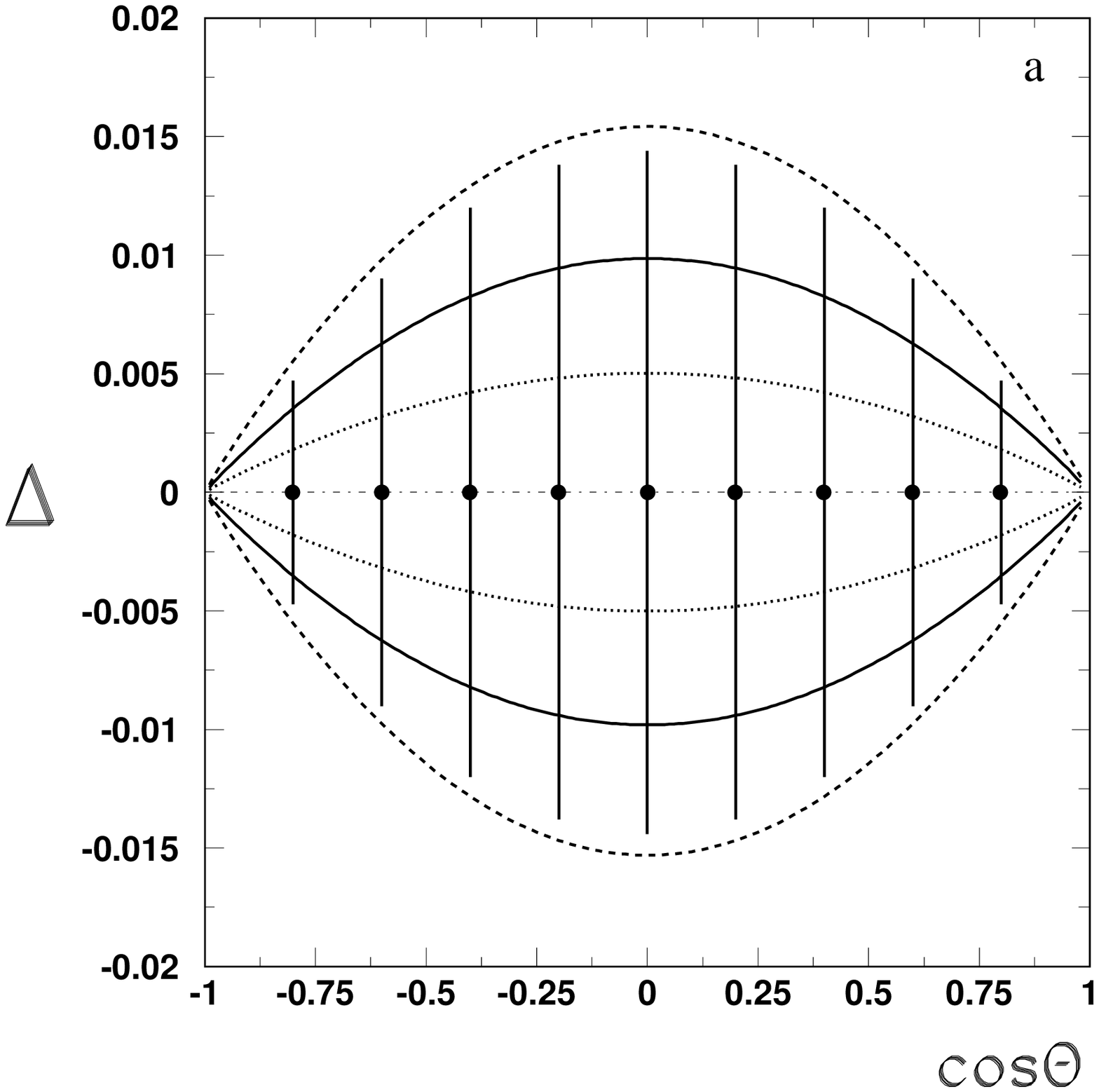}}
 \mbox{\epsfysize=8.5cm\epsffile{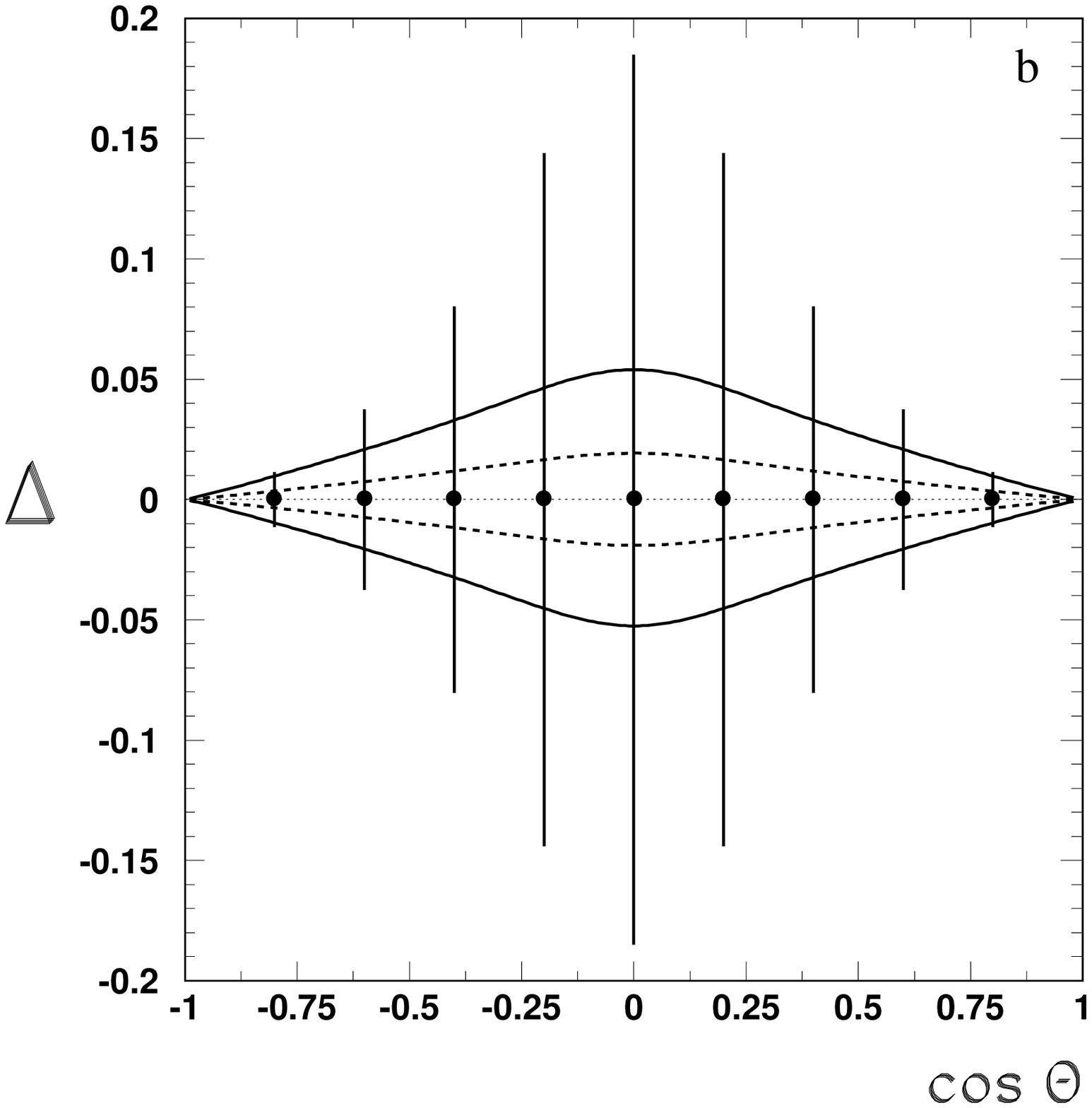}}}
\end{picture}
\vspace*{-5mm}
\caption{
The angular distributions of relative deviations from SM predictions: 
(a) $\Delta(\sigma_{\rm RR})$ for $\Lambda_{\rm RR}$=40 TeV 
(dashed line), 50 TeV (solid line) and 70 TeV (dotted line); (b)  
$\Delta(\sigma_{\rm LR})$ for $\Lambda_{\rm LR}$=30 TeV (solid line) and
50 TeV (dashed line). The curves above (below) the horizontal line 
correspond to negative (positive) interference between contact
interaction and SM amplitude. The error bars show the expected statistical 
relevant uncertainty at $\Lumint{(e^-e^-)}$ same as in Fig.~4.}
\end{center}
\end{figure}
\begin{figure}[ht]
\refstepcounter{figure}
\label{Fig6}
\addtocounter{figure}{-1}
\begin{center}
\setlength{\unitlength}{1cm}
\begin{picture}(12,7.5)
\put(-3.,0.0)
{\mbox{\epsfysize=8.5cm\epsffile{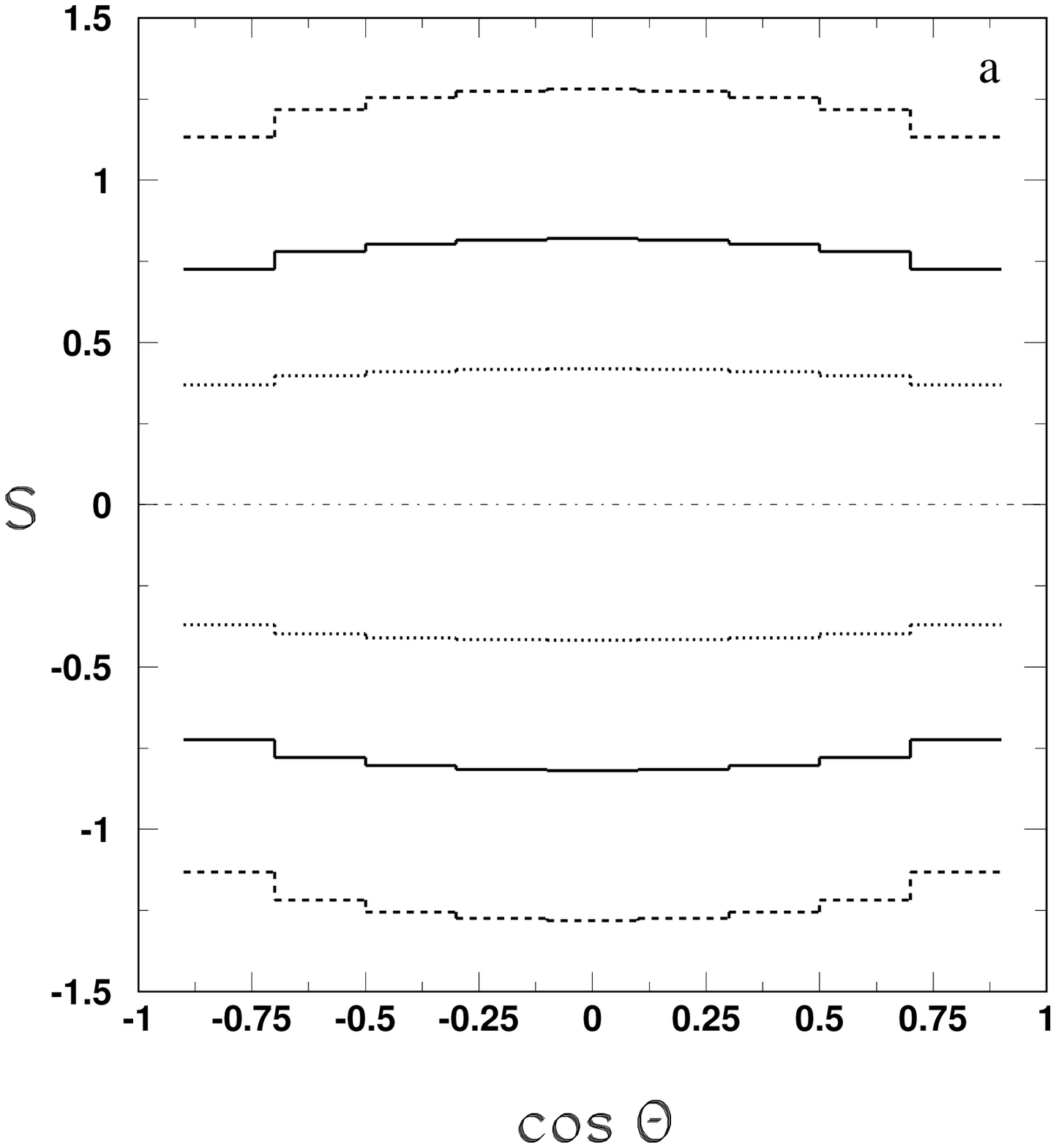}}
 \mbox{\epsfysize=8.5cm\epsffile{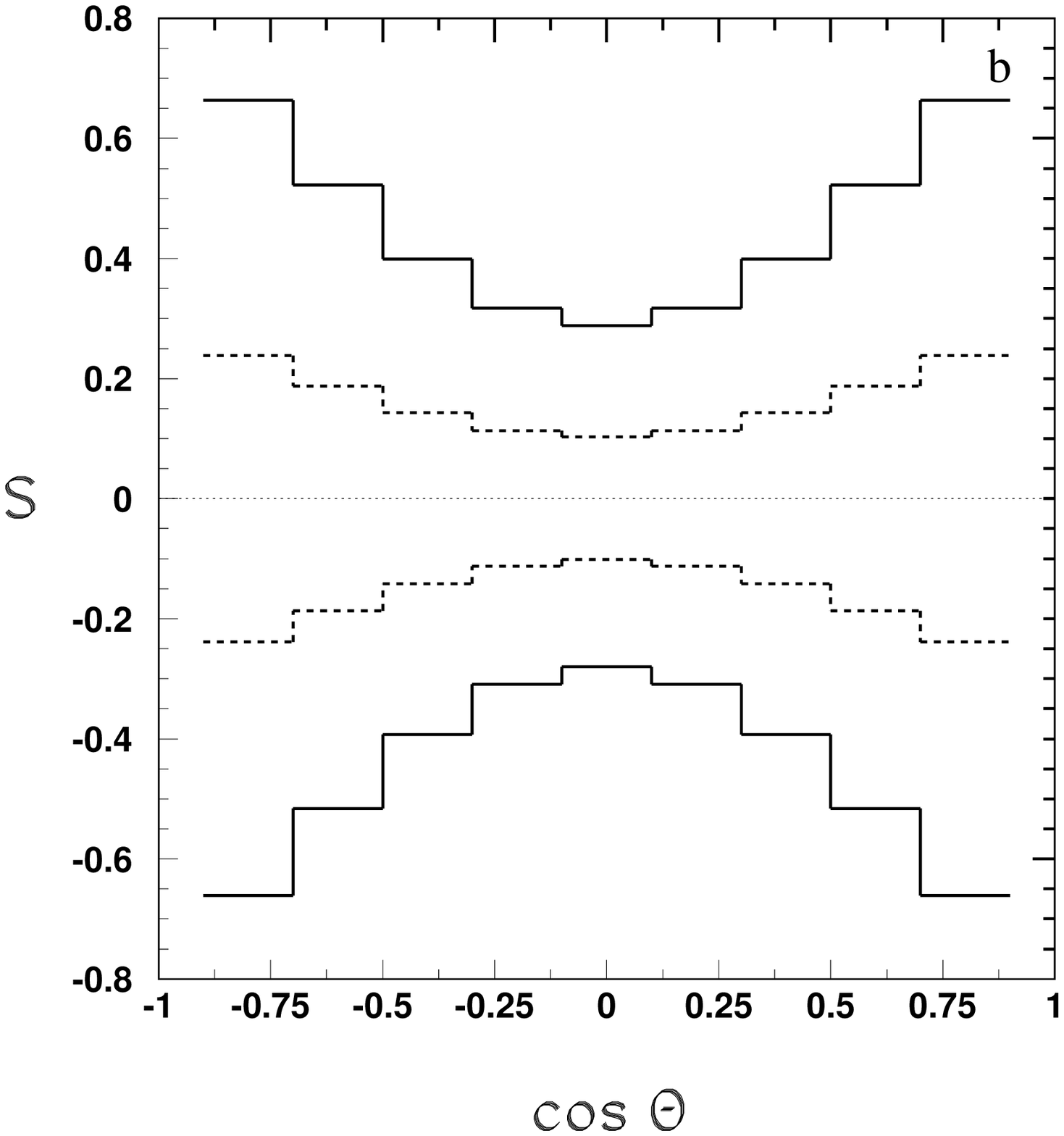}}}
\end{picture}
\vspace*{-5mm}
\caption{
(a) Statistical significance ${\cal S}(\sigma_{\rm RR})$ as a function
of $\cos\theta$ at 
$\Lambda_{\rm RR}$=40 TeV (dashed line), 50 TeV (solid line) and 70
TeV (dotted line); 
(b) Statistical significance ${\cal S}(\sigma_{\rm LR})$ 
as a function of $\cos\theta$ at 
$\Lambda_{\rm LR}$=30 TeV (solid line) and 50 TeV (dashed line).
Here, all inputs are the same as in Fig.~4.
}
\end{center}
\end{figure}
Relative deviations of $\sigma_{\rm RR}$, $\sigma_{\rm LL}$ and 
$\sigma_{\rm LR}$ from the SM model due to the contact interactions 
can be defined in analogy to Eq.~(\ref{relat}). In Fig.~5 we show  
the angular distribution of the deviations $\Delta(\sigma_{\rm RR})$ and 
$\Delta(\sigma_{\rm LR})$, for the values of $\Lumint{(e^-e^-)}$ and 
$\Lambda_{ij}$ indicated in the caption, and with the SM predictions 
evaluated in the same effective Born approximation used in Fig.~4. Such 
deviations are compared to the expected statistical uncertainties represented 
by the vertical bars. The indication of Fig.~5, the analogue of Fig.~2 
for Bhabha scattering, is that, in M{\o}ller scattering, the sensitivity of 
$\sigma_{\rm RR}$ to the related contact parameter $\epsilon_{\rm RR}$ is 
almost flat in $\cos\theta$ leading to high sensitivity to $\epsilon_{\rm RR}$ 
(the same occurs for $\sigma_{\rm LL}$ and $\epsilon_{\rm LL}$). 
Conversely, maximal sensitivity to 
$\epsilon_{\rm LR}$ is obtained in the forward and backward regions where 
the expected statistical uncertainties become smaller. The corresponding 
behaviour of the statistical significance, defined as the ratio 
between deviations and uncertainties for each bin, are shown in Fig.~6, the 
analogue of Fig.~3.    
\par 
We now proceed to the estimate of the constraints on the contact 
interaction couplings from the two processes. 
\section{Numerical analysis and constraints on CI couplings}
We start by assessing the sensivity of Bhabha scattering to the 
compositeness scale. To this purpose, we assume the data to be 
well-described by the SM predictions ($\epsilon_{ij}=0$), {\it i.e.}, 
that no deviation is observed within the foreseen experimental accuracy, 
and perform a $\chi^2$ analysis of the $\cos\theta$ angular distribution. 
For each of the observable cross sections, the $\chi^2$ distribution is 
defined as the sum over the above mentioned nine equal-size $\cos\theta$ 
bins introduced in Sect.~2:
\begin{equation}
\chi^2({\cal O})=
\sum_{\rm bins}\left(\frac{\Delta({\cal O})^{\rm bin}}
{\delta{\cal O}^{\rm bin}}\right)^2\,  =\, 
\sum_{\rm bins}\left[{\cal S}({\cal O})^{\rm bin}\right]^2,
\label{chi}
\end{equation}
where ${\cal O}=\sigma_{\rm L}$, $\sigma_{\rm R}$, $\sigma_{{\rm LR},t}$ 
and $\sigma^{\rm bin}\equiv\int_{\rm bin}(\dd\sigma/\dd z)\dd z$. In 
Eq.~(\ref{chi}), $\Delta({\cal O})$ represents the relative deviation from the 
SM prediction defined in Eq.~(\ref{relat}), and $\delta{\cal O}$ is the 
expected experimental relative uncertainty, that combines the statistical 
and the systematic one.
\par
In order to achieve comparable accuracy in experimental measurements and
theoretical predictions, radiative corrections to Bhabha and M{\o}ller 
scatterings have to be taken into account \cite{denner}. 
In practice, initial state 
radiation is by far the most relevant part of the QED modifications
\cite{nicrosini}. The method that we shall follow to evaluate the
effects of the QED radiation for large-angle Bhabha scattering  is the one 
that uses the so called structure function approach \cite{nicrosini, montagna}
where soft and hard photon emission is taken into account.
As to M{\o}ller scattering, the QED corrections to polarized cross section
will be evaluated by means of the FORTRAN code 
MOLLERAD \cite{shumeiko1, shumeiko2}, adapted to the present discussion, 
with $m_{\rm top}=175$~GeV and $m_H=120$~GeV. 
\par
Concerning the numerical inputs and assumptions used in the estimate of 
$\delta\cal O$, to assess the role of statistics we vary 
$\Lumint{(e^+e^-)}$ from $50$ to $500\ \mbox{fb}^{-1}$ (a third of the 
total running time for each polarization configuration of Eq.~(\ref{cross4})). 
As for the systematic uncertainty, we take $\delta\Lumint/\Lumint=0.5\%$, 
$\delta\epsilon/\epsilon=0.5\%$ and, regarding the electron and positron 
degrees of polarization, $\delta P_1/P_1=\delta P_2/P_2=0.5\ \%$.
\par 
As a criterion to constrain the values of the contact interaction
parameters allowed by the non-observation of the corresponding deviations, 
we impose $\chi^2<\chi^2_{\rm CL}$, where the actual
value of $\chi^2_{\rm CL}$ specifies the desired `confidence' level.
We take the values $\chi^2_{\rm CL}=$3.84 and 5.99 for 95\% C.L. for 
a one- and a two-parameter fit, respectively.
\par 
We begin the presentation of the numerical results from the consideration 
of $\sigma_{\rm L}$ and $\sigma_{R}$. Since these cross sections 
simultaneously depend on the pairs of independent CI couplings 
$(\epsilon_{\rm LL},\epsilon_{\rm LR})$ and 
$(\epsilon_{\rm RR},\epsilon_{\rm LR})$ a two-parameter analysis is needed 
in these cases. The 95\% CL allowed areas are represented by the elliptical 
contours around $\epsilon_{\rm LL}=\epsilon_{\rm RR}=\epsilon_{\rm LR}=0$, 
depicted in Figs.~7a,b. The maximum reachable values of $\Lambda_{\rm RR}$ 
and $\Lambda_{\rm LL}$ correspond to the minimum of the lower branches of 
the curves in these figures. 
\begin{figure}[htb]
\refstepcounter{figure}
\label{Fig7}
\addtocounter{figure}{-1}
\begin{center}
\setlength{\unitlength}{1cm}
\begin{picture}(12,8.)
\put(-3.,0.0)
{\mbox{\epsfysize=8.5cm\epsffile{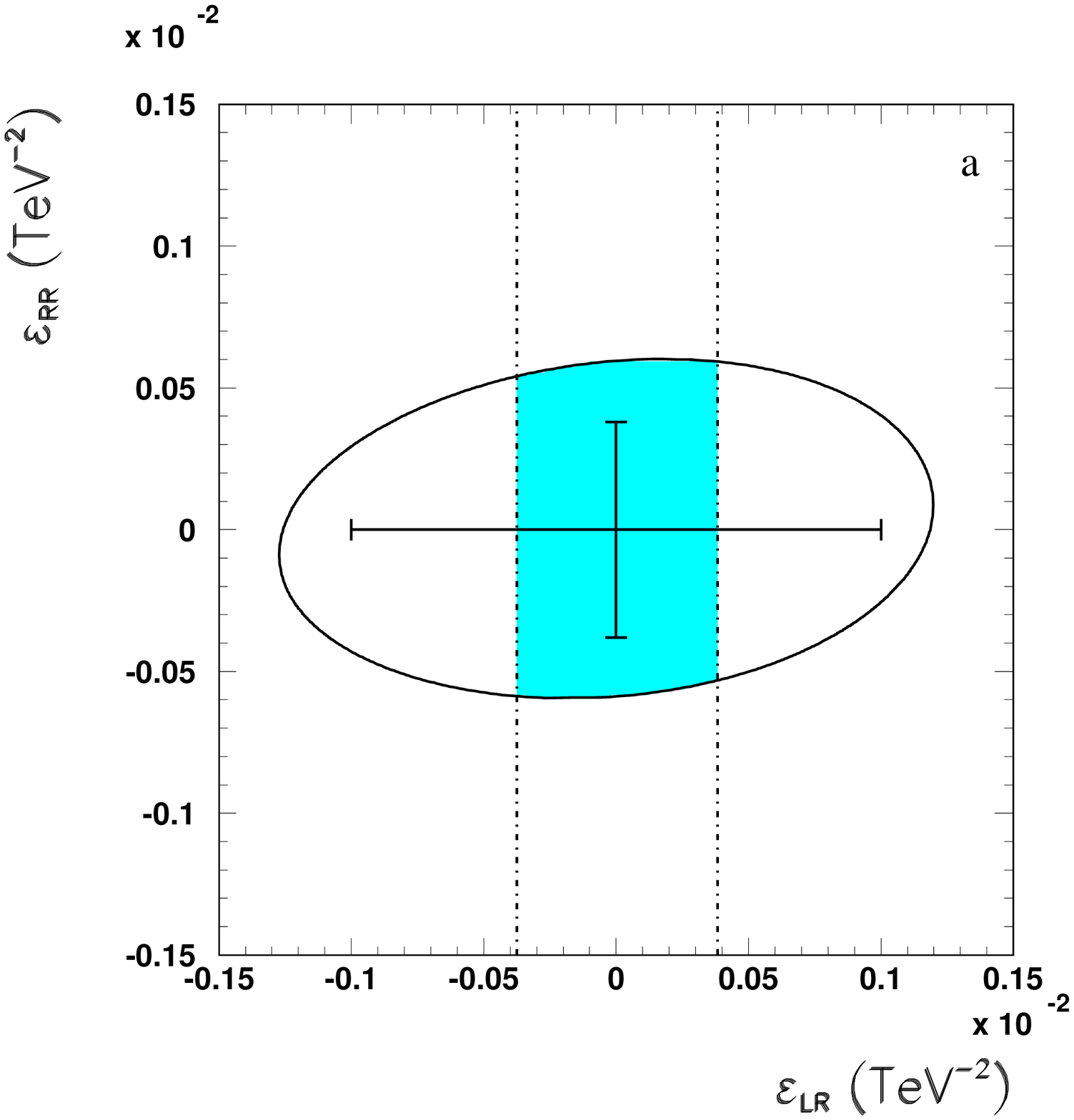}}
 \mbox{\epsfysize=8.5cm\epsffile{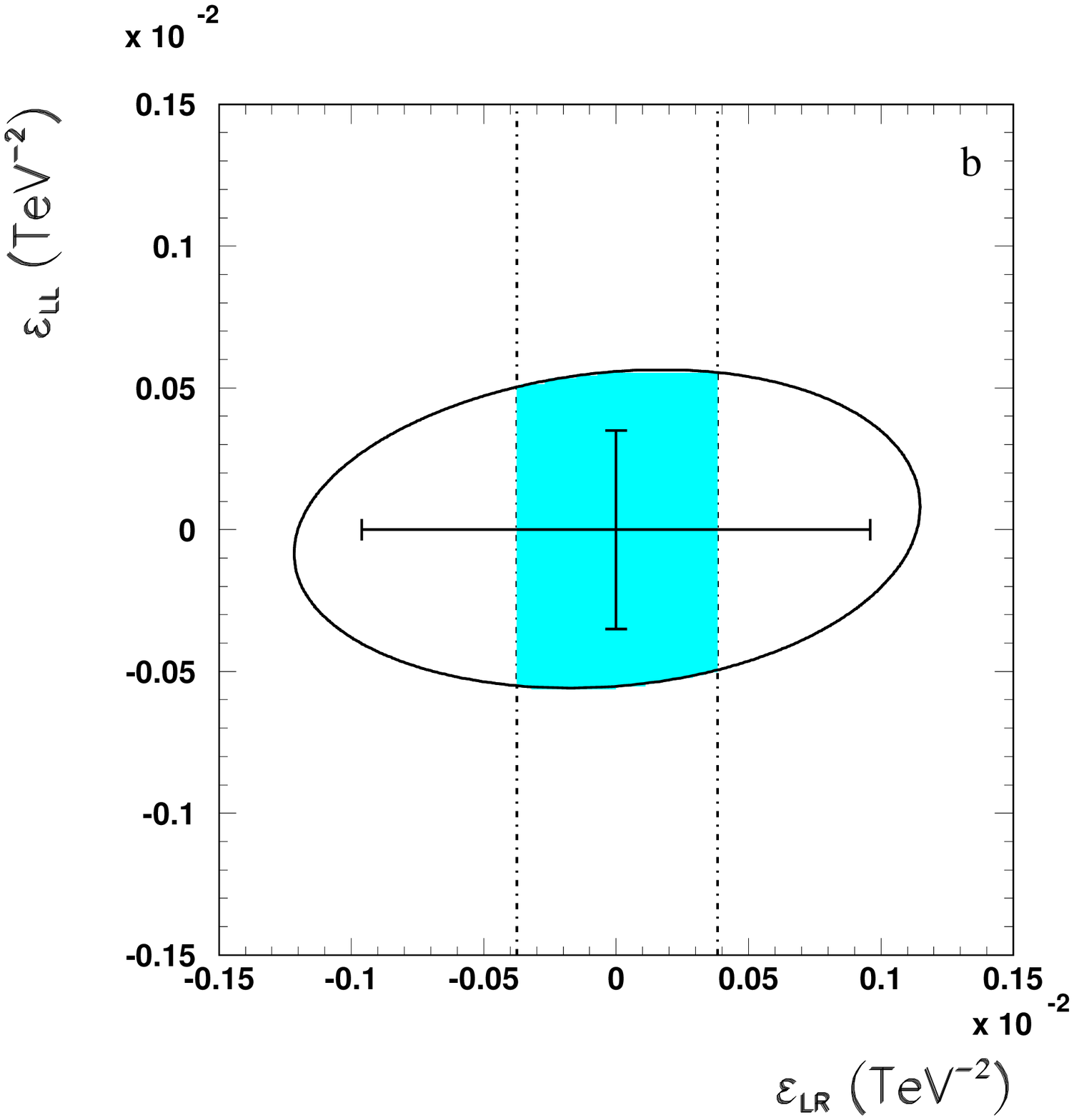}}}
\end{picture}
\vspace*{-3mm}
\caption{
Allowed areas (ellipses) at 95\% C.L. on electron contact 
interaction parameters
in the planes ($\epsilon_{\rm LR},\epsilon_{\rm RR}$) and 
($\epsilon_{\rm LR},\epsilon_{\rm LL}$), 
obtained from $\sigma_{\rm R}$ (a) and $\sigma_{\rm L}$ (b), respectively,
at $\sqrt{s}=500$ GeV, $\Lumint(e^+e^-)=50\ \mbox{fb}^{-1}$,
$\vert P^-\vert=0.8$ and $\vert P^+\vert=0.6$.
Vertical dashed curves indicate the allowed range for 
$\epsilon_{\rm LR}$ obtained from $\sigma_{{\rm LR},t}$.}
\end{center}
\end{figure}
\par 
Turning to $\epsilon_{\rm LR}$, the relevant cross section 
$\sigma_{{\rm LR},t}$ depends only on that parameter, see 
Eqs.~(\ref{helsig}) and (\ref{helamp}), so that the 
corresponding constraints are determined from a one-parameter fit (with 
the lower value of $\chi^2_{\rm CL}$). The model-independent, discovery 
reach expected at the Linear Collider for the corresponding mass scale 
$\Lambda_{\rm LR}$ is represented, as a function of the integrated 
luminosity ${\Lumint}$, in Fig.~8. As expected, the highest luminosity 
determines the strongest constraints on the CI couplings.\footnote{Such 
increase with luminosity is somewhat slower than expected from the scaling 
law $\Lambda\sim\left(s\Lumint\right)^{1/4}$ \cite{barklow}, since with our 
input choice the effect of the systematic uncertainties can compete with the 
statistical one.}  
\par 
The 95\% CL bounds on $\epsilon_{\rm LR}$ can be reported in 
Figs.~7a,b to narrow the constraints on $\epsilon_{\rm RR}$ and 
$\epsilon_{\rm LL}$, respectively. They are represented by the vertical 
lines there, so that the final allowed regions, at the 95\% CL, are the 
shaded ones. Fig.~8 dramatically shows the really high sensitivity of 
$\sigma_{{\rm LR},t}$, such that the discovery limits on 
$\Lambda_{\rm LR}$ are the highest, compared to the $\Lambda_{\rm RR}$ and 
$\Lambda_{\rm LL}$ case.
\begin{figure}[thb]
\refstepcounter{figure}
\label{Fig8}
\addtocounter{figure}{-1}
\begin{center}
\setlength{\unitlength}{1cm}
\begin{picture}(8.,8.)
\put(-1.0,-1.5){
\mbox{\epsfysize=10.0cm\epsffile{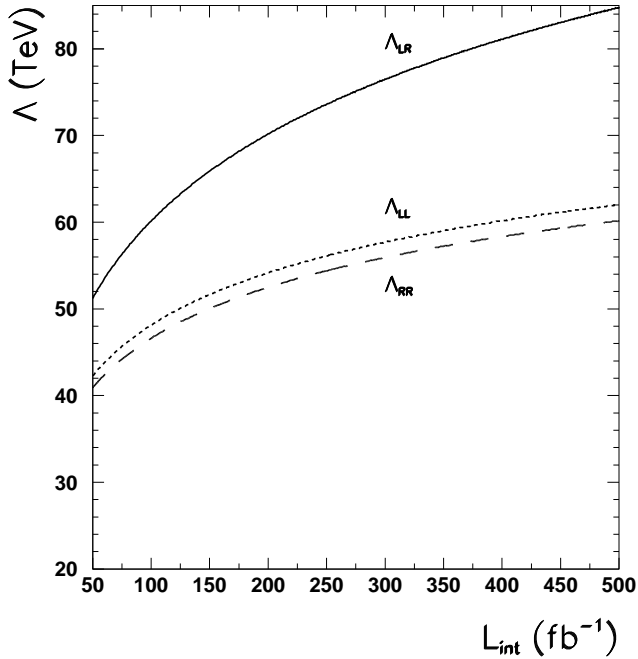}}}
\end{picture}
\vspace*{13mm}
\caption{
Reach in $\Lambda$ at 95\% C.L. {\it vs.} integrated 
luminosity ${\Lumint}(e^+e^-)$ obtained from the model-independent analysis
for $e^++e^-\to e^++e^-$ at $E_{\rm c.m.}=0.5$~TeV,
$\vert P^-\vert=0.8$ and $\vert P^+\vert=0.6$, $\Lambda_{\rm LR}$ (solid
line),
$\Lambda_{\rm RR}$ (dashed line), $\Lambda_{\rm LL}$ (dotted line).
}
\end{center}
\end{figure}
\par 
The crosses in Fig.~7a,b represent the model-dependent constraints 
obtainable by taking only one non-zero parameter at a time, instead 
of two simultaneously non-zero and independent as in the analysis 
presented above. The arms of the crosses refer to integrated luminosity 
$\Lumint=50\ \mbox{fb}^{-1}$. One can note from Figs.~7a,b that the 
`single-parameter' constraints on the individual CI parameters 
$\epsilon_{\rm RR}$ and $\epsilon_{\rm LL}$ are numerically more stringent, 
as compared to the model-independent ones. Essentially, this is a reflection 
of the smaller critical value of $\chi^2$, $\chi^2_{\rm crit}=3.84$, 
corresponding to 95\% C.L. with a {\it one-parameter} fit. 
\par
The procedure, and the criteria, to derive numerical constraints from the 
M{\o}ller process are quite similar, the outstanding difference being that, 
in this case, {\it each} measurable cross section in (\ref{observmol}) 
depends on a single contact interaction parameter, so that complete 
disentangling of $\epsilon$'s is directly obtained and the smaller 
$\chi^2_{\rm CL}=3.84$, relevant to one-parameter cases, applies. Certainly, 
this is an advantage if one wants to perform a model-independent analysis of 
electron contact interactions. Also, substantially higher longitudinal 
polarization should be attainable for electron beams than for positron ones, 
for a given luminosity. On the other side, there is the penalty of the lower 
luminosity expected in the $e^-e^-$ mode, depressing the sensitivity, and, as 
previously stated, in our examples we have assumed a third of the luminosity 
in the $e^+e^-$ mode. The lower bounds on $\Lambda$'s, derived under these 
conditions, are shown as a function of the integrated luminosity in Fig.~9.    
\begin{figure}[thb]
\refstepcounter{figure}
\label{Fig9}
\addtocounter{figure}{-1}
\begin{center}
\setlength{\unitlength}{1cm}
\begin{picture}(8.,8.)
\put(-1.0,-1.5){
\mbox{\epsfysize=10.0cm\epsffile{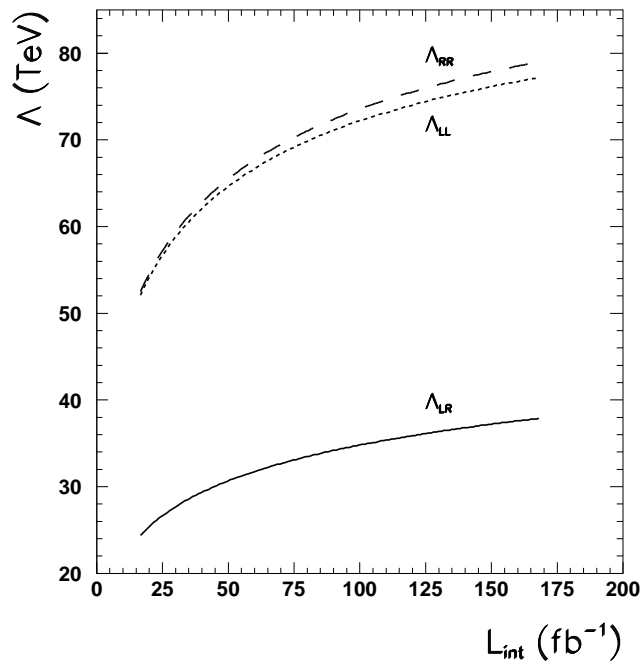}}}
\end{picture}
\vspace*{12mm}
\caption{
Reach in $\Lambda$ at 95\% C.L. {\it vs.} integrated 
luminosity ${\Lumint}(e^-e^-)$ obtained from the model-independent analysis
for $e^-+e^-\to e^-+e^-$ at $E_{\rm c.m.}=0.5$~TeV,
$\vert P_1^-\vert=\vert P_2^-\vert=0.8$, $\Lambda_{\rm LR}$ (solid line),
$\Lambda_{\rm RR}$ (dashed line), $\Lambda_{\rm LL}$ (dotted line).
}
\end{center}
\end{figure}
\section{Concluding remarks}
In the previous sections we have derived limits on the electron contact 
interactions by simultaneously considering Bhabha scattering and M{\o}ller 
scattering at a Linear Collider with longitudinally polarized beams, using a  
model-independent analysis that allows to simultaneously account for all 
independent couplings as non-vanishing free parameters. The analysis is based  
on the definition of measurable polarized differential cross sections that 
allows to derive: {\it i)} from Bhabha scattering, separate bounds on 
$\epsilon_{\rm LR}$ and in the planes $(\epsilon_{\rm LL},\epsilon_{\rm LR})$ 
and $(\epsilon_{\rm RR},\epsilon_{\rm LR})$; {\it ii)} from M{\o}ller 
scattering,
completely individual bounds on $\epsilon_{\rm LL}$, $\epsilon_{\rm RR}$ and 
$\epsilon_{\rm LR}$. Numerical results for the lower bounds on the 
corresponding range in the relevant mass scales $\Lambda_{ij}$, depending 
on the luminosity, are shown in Figs.~8 and 9, and are summarized in Table~1. 
Essentially, for c.m. energy $\sqrt s = 500$ GeV and reasonable assumptions 
on luminosities, polarizations and their relative uncertainties, 
in the Bhabha mode the bounds vary from 41 to 62 TeV for the LL and RR cases, 
and from 51 to 85 TeV for the LR coupling. In the M{\o}ller mode, the 
bounds vary from 52 to 79 TeV for the LL and RR cases, and from 24 to 
38 for the LR coupling (notice the reduced luminosity input in this case). 
Therefore, 
one can conclude that the two processes are complementary as far as the 
sensitivity to the individual couplings in a model-independent data 
analysis is concerned: the sensivity of Bhabha scattering to 
$\Lambda_{\rm LR}$ is 
dramatically higher, while M{\o}ller scattering is the most sensitive to 
$\Lambda_{\rm LL}$ and $\Lambda_{\rm RR}$. Basically, for the inputs 
used in Figs.~8 and 9, the ratio of the maximal sensitivities to $\Lambda$ 
of the two processes has the qualitative behaviour: 
\begin{equation}
\frac{\Lambda^{e^-e^-}_{\rm RR}}{\Lambda^{e^+e^-}_{\rm LR}}\approx 
\left(2\, \frac{\Lumint{(e^-e^-)}}{\Lumint{(e^+e^-)}}\right)^{1/4}
\approx 0.9\hskip 3pt .
\end{equation}  
\begin{table}[th]
\centering
\caption{
Reach in $\Lambda_{ij}$ at 95\% C.L., from the
model-independent analysis performed 
for $e^+e^-\to\mu^+\mu^-$ and $e^+e^-$, 
at $E_{\rm c.m.}=0.5$~TeV, $\Lumint=50\,\mbox{fb}^{-1}$ and
$500\,\mbox{fb}^{-1}$, $\vert P^-\vert=0.8$ and 
$\vert P^+\vert=0.6$.
}
\medskip
{\renewcommand{\arraystretch}{1.2}
\begin{tabular}{|c|c|c|c|c|c|}
\hline
${\rm process}$ & ${\Lumint}$ & $\Lambda_{\rm LL}$ & $\Lambda_{\rm RR}$ &
$\Lambda_{\rm LR}$ & $\Lambda_{\rm RL}$ \\
&$\mbox{fb}^{-1}$& TeV & TeV &TeV  & TeV \\
\hline
\cline{2-6}
 & 50 & $ 35 $ & $ 35 $ & $ 31 $ & $ 31 $
\\
\cline{2-6}
$e^+e^-\to\mu^+\mu^-$
 & 500 &$ 47 $ & $ 49 $ & $ 51 $ & $ 52 $ 
\\ \hline
\cline{2-6}
 & 50 & $ 42 $ & $ 41 $ & $ 51 $ &
\\
\cline{2-6}
$e^+e^-\to e^+ e^-$
 & 500 & $ 62 $ & $ 60 $ & $ 85 $ & 
\\ \hline
\cline{2-6}
 & 50/3 & $ 52 $ & $ 53 $ & $ 24 $ &
\\
\cline{2-6}
$e^-e^-\to e^- e^-$
 & 500/3 & $ 77 $ & $ 79 $ & $ 38 $ &  
\\ \hline
\end{tabular}
} 
\label{tab:table-1}
\end{table}
\par 
All this shows the benefits of initial beams longitudinal polarization, 
that allows, by measuring suitable combinations of polarized cross sections, 
to directly disentangle the individual couplings. Indeed, as previously
observed, in general without polarization only correlations among contact 
interaction parameters, rather than finite allowed regions, could be derived 
and consequently, in the unpolarized case, only a one-parameter analysis, 
relating to a specific model, can be performed.  
\par 
In Table~1 we have also reported the numerical results relevant to the 
annihilation into muon pairs, derived from a similar analysis
\cite{babich2001}.  
\par 
As an example of application of the obtained results to a possible 
source of contact interactions, we may consider the sneutrino 
parameters (mass $m_{\tilde\nu}$ and Yukawa coupling $\lambda$)  
envisaged by supersymmetric theories with ${\cal R}$-parity breaking.
In this case, sneutrino exchange affects
only those helicity amplitudes with non-diagonal chiral indices, so that  
$\Lambda_{\rm LR}$ is the relevant mass scale \cite{rizzo1, zerwas}. 
Qualitatively, without entering into a detailed and more complex analysis, 
one can expect typical bounds on 
$m_{\tilde\nu}/\lambda\sim\Lambda_{\rm LR}/\sqrt{8\pi}\simeq 10$ to 17 TeV 
corresponding to $\Lambda_{\rm LR}\approx 51$ TeV and 85 TeV (Fig.~8) at
$\Lumint=50\ \mbox{fb}^{-1}$ and $500\ \mbox{fb}^{-1}$, respectively.
 
\medskip
\leftline{\bf Acknowledgements}
\par\noindent
This research has been partially supported by MIUR (Italian Ministry of 
University and Research) and by funds of the University of Trieste.
\goodbreak


\end{document}